\documentclass[aps,reprint,prd,nofootinbib,noshowkeys,longbibliography,floatfix,superscriptaddress,preprintnumbers]{revtex4-1}
\usepackage[utf8]{inputenc} 
\usepackage{hyperref} 
\usepackage[left=3cm,right=3cm,top=3cm,bottom=3cm]{geometry} 
\usepackage{amsmath} 
\usepackage{amsfonts}
\usepackage{amsthm}
\usepackage{amssymb}
\usepackage{graphicx} 
\usepackage{slashed} 
\usepackage{color} 
\usepackage[sort&compress]{natbib}

\graphicspath{{./}{./pics/}}

\newcommand{\mr}[1]{\mathrm{#1}}
\newcommand{\mb}[1]{\mathbf{#1}}
\newcommand{\mc}[1]{\mathcal{#1}}
\newcommand{\ti}[1]{\tilde{#1}}
\newcommand{\Aext}{A^{\mathrm{ext}}_\mu}
\newcommand{\Aextdual}{\tilde{A}^{\mathrm{ext}}_\mu}
\newcommand{\Fext}{F^{\mathrm{ext}}_{\mu\nu}}
\newcommand{\bmax}{b_{\mr{max}}}
\newcommand{\kp}{\xi} 
\newcommand{\rcl}{r_{\mr{cl}}} 

\begin{document}

\title{Towards Schwinger production of magnetic monopoles in heavy-ion collisions}
\date{February 13, 2019}
\author{Oliver Gould}
\email{oliver.gould@helsinki.fi}
\affiliation{Helsinki Institute of Physics, University of Helsinki, FI-00014, Finland}
\author{David L.-J. Ho}
\email{d.ho17@imperial.ac.uk}
\affiliation{Department of Physics, Imperial College London, SW7 2AZ, UK}
\author{Arttu Rajantie}
\email{a.rajantie@imperial.ac.uk}
\affiliation{Department of Physics, Imperial College London, SW7 2AZ, UK}
\preprint{IMPERIAL-TP-2019-DH-01}
\preprint{HIP-2019-2/TH}

\begin{abstract}
Magnetic monopoles may be produced by the Schwinger effect in the strong magnetic fields of peripheral heavy-ion collisions.
We review the form of the electromagnetic fields in such collisions and calculate from first principles the cross section for monopole pair production. Using the worldline instanton method, we work to all orders in the magnetic charge, and hence are not hampered by the breakdown of perturbation theory.
Our result depends on the spacetime inhomogeneity through a single dimensionless parameter, the Keldysh parameter, which is independent of collision energy for a given monopole mass.
For realistic heavy-ion collisions, the computational cost of the calculation becomes prohibitive and the finite size of the monopoles needs to be taken into account, and therefore our current results are not applicable to them.
Nonetheless, our results show that the spacetime dependence enhances the production cross section and would therefore lead to stronger monopole mass bounds than in the constant-field case.
\end{abstract}

\maketitle

\section{Introduction}\label{sec:introduction}

Magnetic monopoles, hypothetical particles with a single magnetic pole, are present in generic classes of theories beyond the Standard Model, and their existence would explain the quantisation of electric charge~\cite{dirac1931quantised,Preskill:1984gd,Polchinski:2003bq}.
Roughly speaking, there are two different types of magnetic monopoles: solitonic and elementary. 
Solitonic monopoles, such as 't Hooft-Polyakov monopoles~\cite{thooft1974magnetic,polyakov1974particle}, which exist in all Grand Unified Theories (GUTs),
are smooth semiclassical solutions of the field equations with a nonzero physical size.   
Their mass is determined by the parameters of the theory, and in typical GUTs, it is very high, above $10^{16}~{\rm GeV}$. There have been attempts to construct theories with lower mass solitonic monopoles~\cite{Cho:1996qd,Dienes:1998vg,Bruemmer:2009ky,Ellis:2016glu,Arunasalam:2018iom,Arai:2018uoy,Mavromatos:2018kcd}, which would bring them within the reach of particle experiments.
In contrast, 
elementary Dirac monopoles~\cite{cabibbo1962quantum,schwinger1966magnetic,zwanziger1970local,blagojevic1985quantum}
appear as fields in the Lagrangian, in the same way as any Standard Model particle, and therefore
the bare particles are pointlike. Their mass is a free parameter, only constrained by experiments.

At large distances the properties of magnetic monopoles depend only on their
mass, $m$, and two discrete parameters, determining their spin, $s$, and magnetic charge, $g$. In principle they may also have an electric charge~\cite{Zwanziger:1968rs,Julia:1975ff,Witten:1979ey}.
In contrast, the short-distance details of magnetic monopoles vary depending on the theory, in particular depending on whether the monopoles are elementary particles or semiclassical solitons.

The possibility of producing magnetic monopoles in particle colliders has been considered extensively \cite{Rajantie:2016paj,PDG:2018PRD} and is currently the focus of a dedicated experiment at the Large Hadron Collider (LHC), MoEDAL \cite{Acharya:2017cio}. In the absence of a positive discovery, these experiments place upper bounds on the monopole production cross section. To turn these into constraints on the theory, one would need a reliable theoretical description of the monopole production process.
It is conventional for experiments to report mass bounds based on the tree-level Drell-Yan cross section
\begin{equation}
\sigma_{DY} = \frac{q_q^2 g^2}{12\pi E^2},
\label{equ:sigmaDrellYan}
\end{equation}
where $q_q$ is the quark electric charge and $E$ is centre-of-mass energy, even though it is known to be inapplicable. The difficulty is that magnetic monopoles are necessarily strongly coupled due to the Dirac quantisation condition \cite{dirac1931quantised}, which inversely relates magnetic and electric charges, $g$ and $e$, by
\begin{equation}
eg=2\pi n,\quad n\in\mathbb{Z}. 
\label{equ:Diraccondition}
\end{equation}
Thus the magnetic fine structure constant is  $g^2/4\pi=\pi/e^2\approx 34\gg 1$, implying that perturbation theory breaks down.

It is believed~\cite{Witten:1979kh,drukier1982monopole} that in proton-proton collisions, the production of solitonic monopoles, such as GUT monopoles and other 't Hooft-Polyakov monopoles, is suppressed by
\begin{equation}
\sigma_{pp\to M\bar{M}}\propto \mr{e}^{-4/\alpha} \sim 10^{-236},\label{eq:suppression}
\end{equation}
independently of collision energy. This overwhelming factor would rule out the production of solitonic magnetic monopoles in proton-proton collisions, even for very high collision energies and luminosities. The suppression can be understood as arising from the large number of degrees of freedom in the final state compared to the initial few-particle state, or from the exponentially small overlap between the hard initial state and the spatially extended final state. Although Eq.~\eqref{eq:suppression} is somewhat conjectural for magnetic monopoles, it has been explicitly demonstrated for the analogous processes of scalar vacuum decay \cite{Kuznetsov:1997az}, semiclassical $(B-L)$ violating processes \cite{Rebbi:1996zx,Bezrukov:2003er,Bezrukov:2003qm} (though there has been some dispute in this case \cite{Tye:2015tva,Qiu:2018wfb}) and scalar soliton production \cite{Demidov:2011dk,Papageorgakis:2014dma,Demidov:2015nea,Demidov:2015bua}.

For elementary Dirac monopoles these arguments do not apply. However, once dressed, elementary monopoles are far from point-like. It has been argued that photon-magnetic monopole interactions are effectively delocalised on the scale of the classical radius, or Thompson scattering length, of the monopole, $\rcl=g^2/(4\pi m)\gg 1/m$ \cite{goebel1970spatial,goldhaber1982monopoles,Coleman:1982cx}. The original argument of Ref.~\cite{goebel1970spatial}, in an S-matrix language, relies on the Thompson formula \cite{low1954scattering,gellmann1954scattering}, the Kramers-Kronig dispersion relations \cite{goldberger1955use} and the optical theorem, all of which are valid beyond the weak coupling expansion. Note that for weakly coupled particles, such as electrons, the classical radius is smaller than the Compton wavelength and hence is dynamically irrelevant. Thus it is the strong coupling of magnetic monopoles which leads to their large effective size.

Such a dressed elementary monopole state will have an exponentially small overlap with any hard state with energy $E\sim m$. As a consequence, one would expect the cross section for elementary monopole production from a hard initial state also to be exponentially suppressed,
\begin{align}
\sigma_{pp\to M\bar{M}}&\propto|\langle M \bar{M} |\hat{\mc{O}}| E \rangle|^2, \nonumber \\
&\sim \bigg|\int \mr{d}x\ \psi_{M\bar{M}}(x)\mc{O}\mr{e}^{-iEx}\bigg|^2, \nonumber \\
&\lesssim \mr{e}^{-2E\rcl}\leq \mr{e}^{-4/\alpha},
\end{align}
where $\hat{\mc{O}}$ is some operator characterising the interaction. As long as $\mc{O}$ is not exponentially large, the exponential suppression should not depend on it. This argument follows that of Landau \cite{landau1932theoryi,landau1932theoryii,landau1948theoryiii,landau1991quantum} (see also Section 7 of Ref.~\cite{Khoze:2018mey} for a recent discussion). In going from the second to the third line we have assumed the monopole-antimonopole state, $\psi_{M\bar{M}}$, to be a smooth function, varying on the scale $r_{\mr{cl}}$ and have used the Riemann-Lebesgue lemma (see also \cite{Papageorgakis:2014dma}). On the third line we have used that $E\geq 2m$ for monopole production to be kinematically possible.
Of course this is not a complete argument, but it means that mass bounds obtained by assuming the tree-level Drell-Yan cross section (\ref{equ:sigmaDrellYan}) may be off by many orders of magnitude.

In heavy-ion collisions, none of these arguments for exponential suppression apply. This is because the fundamental process of magnetic monopole pair production does not proceed from a hard initial state with a small number of degrees of freedom. Instead, pair production proceeds by the quantum-mechanical decay of a classically-occupied electromagnetic field, the Schwinger mechanism \cite{sauter1931verhalten,heisenberg1936folgerungen,schwinger1951gauge,affleck1981monopole,affleck1981pair}. This nonperturbative process cannot be reduced to a sum over processes involving small, finite numbers of photons.

The magnetic fields present in heavy-ion collisions are the strongest in the known universe \cite{huang2015electromagnetic}.
Stronger fields give a greater probability of pair production, so heavy-ion collisions provide the most promising terrestrial possibility of producing magnetic monopoles. A reliable computation of the production cross section for magnetic monopoles in these collisions is thus of high experimental and theoretical interest. If this can be achieved, then, at particle colliders such as the LHC, it will be possible to confirm or rule out the existence of magnetic monopoles with masses in a certain, computable range.

A comprehensive review of the electromagnetic fields in heavy-ion collisions can be found in Ref.~\cite{huang2015electromagnetic}. Two electron-stripped ions (commonly lead, gold or uranium) travel towards each other at highly relativistic speeds, generating strong electromagnetic fields. After the collision, a quark-gluon plasma (QGP) is believed to form within a time \(\tau_0 \sim 0.2\text{-}0.6 \, \mathrm{fm}/c\) \cite{busza2018heavy} (at LHC energies). However, for ultrarelativistic collision energies the magnetic field is expected to decrease in strength significantly from its peak magnitude before QGP formation: for TeV collisions at LHC the decay timescale for the magnetic field is \(O(10^{-3} \, \mr{fm}/c)\). As a consequence, we do not expect the QGP to have a significant effect during the time which monopoles are most likely to be produced, so we do not include this in our analysis. Furthermore as the timescales we consider occur well before thermalisation (which is necessarily after the time of QGP formation), we do not include finite-temperature effects.

Bounds on monopole masses that do not rely on perturbative techniques are currently scarce and lenient. The earliest reliable bounds arise from the expectation that sufficiently light monopoles would be produced thermally during reheating \cite{turner1982thermal,collins1984thermal,lindblom1985thermal}. Using experimental bounds on monopole flux \cite{parker1970origin,turner1982magnetic,turner1993extension} and noting that the universe during reheating must have been hotter than during Big Bang Nucleosynthesis, the mass bound \(m \gtrsim 0.45 \, \mr{GeV}\) can be obtained. In Ref.~\cite{gould2018mass} somewhat stronger bounds were obtained by considering Schwinger production, giving a lower bound of \(O(1 \, \mr{GeV})\) dependent on the monopole charge.

The most stringent mass bounds have been obtained by considering Schwinger production in relatively low-energy ($\sqrt{s_{NN}}\approx 17~\mr{GeV}$) heavy-ion collisions at the Super Proton Synchrotron (SPS) \cite{he1997search,gould2018mass}. A magnetic field constant in both space and time was assumed, in addition to a finite temperature. In such collisions QGP formation occurs over a timescale comparable to the decay time of the fields, so it was argued there that thermal affects should be taken into account. For LHC collisions, however, these assumptions are not expected to be valid.

In this paper we relax the constant-field assumption and consider monopole production in the inhomogeneous electromagnetic fields in ultrarelativistic heavy-ion collisions. We present approximate analytical expressions for the fields that fit well to direct numerical integrations for ultrarelativistic collisions. For reasons outlined above, we neglect thermal effects.

Schwinger production in inhomogeneous fields at weak coupling has been subject to previous study \cite{brezin1970pair,Nikishov:1970br,Popov:1971ff,popov1972pair,Marinov1977electron,dunne2005worldline,dunne2006worldline,kim2006schwinger} --- results indicate that spatial inhomogeneity tends to suppress production whilst time dependence enhances it. We argue that due to the form of the magnetic fields in heavy-ion collisions, when considering monopole production, the effects of time dependence dominate over those of the spatial dependence, leading to strongly enhanced production over the constant field case. Furthermore, our results suggest that the effect of the time dependence on the functional form of the production probability is \emph{independent} of collision energy for a given monopole mass.

A careful consideration of the validity of our approximations shows that the parameter regions in which our results are valid are unfortunately unobtainable in real heavy-ion collisions. This is due, in part, to the particular values of charges and radii of stable nuclei that happen to exist in nature. In an alternative universe where significantly higher nuclear charges are possible, our approximations are sound and predict rather large production cross sections for magnetic monopoles. If we naively extrapolate our results beyond their region of applicability, to leading order in an expansion in monopole self-interactions they suggest that one could produce monopoles in the hundreds of GeV mass range at the LHC (see Fig.~\ref{fig:crossSectionPlot}). Further, higher-order corrections only seem to enhance the cross section. Despite the breakdown of our approximations, we provide answers to important questions regarding the effect of field inhomogeneity on monopole production, and reveal the next steps required to obtain reliable mass bounds in the LHC era.

The paper is organised as follows. In Section \ref{sec:general_approach} we outline our general approach to the computation of the monopole production cross section, briefly reviewing the worldline instanton formalism as applied to magnetic monopoles. In Section \ref{sec:em_fields} we calculate the electromagnetic fields in ultrarelativistic heavy ion-collisions and find a simple fit to their functional form. Armed with this, we compute the worldline instanton in Section \ref{sec:worldlineInstanton}, analytically and numerically in certain regions of parameter space. Appendix \ref{appendix:discretisation} gives details of our numerical discretisation. In Section \ref{sec:consequences} we discuss the consequences of our results for magnetic monopole searches and in Section \ref{sec:conclusions} we conclude.

Throughout we use units such that $c=\hbar=\epsilon_0=k_B=1$.

\section{General approach}\label{sec:general_approach}

If magnetic monopoles exist, then magnetic fields can decay into magnetic monopole pairs \cite{affleck1981monopole}. This is the electromagnetic dual of Sauter-Schwinger pair production \cite{sauter1931verhalten,heisenberg1936folgerungen,schwinger1951gauge}.

Schwinger pair production can be formulated as a vacuum decay process. In this case the so-called false vacuum, $|\Omega\rangle$, is the vacuum state in the absence of the external field. It contains no charged particles. The probability of the decay of this state is given by
\begin{equation}
 P=1-|\langle \Omega | \hat{S} | \Omega \rangle|^2=1-\mathrm{e}^{2\mr{Im}\left( iW\right)},
\end{equation}
where $\hat{S}$ is the S-matrix including the external field and $W$ is defined by $\mr{e}^{iW}:=\langle \Omega | \hat{S} | \Omega \rangle$. For slow vacuum decays, when the decay rate is much slower than other relevant timescales, the calculation of $P$ can be formulated in Euclidean time \cite{langer1967theory,langer1969statistical,coleman1977fate,callan1977fate}. In this case
\begin{equation}
 P\approx 2\mr{Im}\left( W_E\right),
\end{equation}
where $W_E$ is defined by $\mr{e}^{-W_E}:=\langle \Omega | \hat{S}_E | \Omega \rangle$ and $\hat{S}_E$ is the Wick rotated ``S-matrix'', again including the external field.

In the following we consider Schwinger pair production to be the only mechanism of pair production. Thus our results provide a lower bound on the true cross section of pair production. The problem then factorises into i) the calculation of the electromagnetic field as a function of the collision parameters and ii) the calculation of the probability to produce magnetic monopoles from a given electromagnetic field.

We will treat the electromagnetic field of the ions as a classical background or external field. We include the effect of quantum photon fluctuations to the pair production process itself, though we do not include contributions from fluctuations inherent to the ions. With this assumption, the cross section for magnetic monopole pair production takes the form,
\begin{equation}
 \frac{\mr{d}\sigma_{M\bar{M}}}{\mr{d}b} = 2\pi b\ P\left(\Fext(\sqrt{s},b)\right), \label{eq:crosssection_def}
\end{equation}
where the factor of $2\pi b$ is simply the geometric differential cross section, and $\Fext(\sqrt{s},b)$ is the classical electromagnetic field. Note that it does not matter whether or not the ions actually collide, as strong electromagnetic fields are also produced by near-misses. We leave determining $\Fext(\sqrt{s},b)$ to Section \ref{sec:em_fields}. For the rest of this section, we will outline the calculation of the probability of pair production for a given electromagnetic field.

Magnetic monopoles couple to to the gauge field which generates the electromagnetic dual field, $\ti{F}^{\mu\nu} := \tfrac{1}{2}\epsilon^{\mu\nu\rho\sigma}F_{\rho\sigma}$, where $\epsilon^{\mu\nu\rho\sigma}$ is the Levi-Civita symbol, $\epsilon^{0123}=1$,
\[
\ti{F}_{\mu\nu} =
\begin{pmatrix}
0 & B_1 & B_2 & B_3 \\
-B_1 & 0 & E_3 & -E_2 \\
-B_2 & -E_3 & 0 & E_1 \\
-B_3 & E_2 & -E_1 & 0 \\
\end{pmatrix}.
\]
The dual gauge field, $\ti{A}_\nu$, satisfies $\ti{F}_{\mu\nu}=\partial_\mu \ti{A}_\nu - \partial_\nu\ti{A}_\mu$, and is simply a rearrangement of the usual two degrees of freedom of the photon field --- it contains no extra degrees of freedom.

We first consider elementary, scalar magnetic monopoles, $\phi$, with charge $g$ and mass $m$. The introduction of the external field, $\Aext$, is achieved by shifting the gauge field in the covariant derivative of $\phi$. The Euclidean Lagrangian for the photon field, $A_\mu$, coupled to spin 0 monopoles is then
\begin{align}
 \mathcal{L}_{s=0}&:=\frac{1}{4}F_{\mu \nu}F_{\mu \nu} + \tilde{D}_\mu \phi ( \tilde{D}_\mu \phi)^* \nonumber \\
 &\qquad + m^2\phi\phi^*+ \frac{\lambda}{4}(\phi\phi^*)^2, \label{eq:sqedlagrangian}
\end{align}
where $F_{\mu \nu}=\partial_\mu A_\nu-\partial_\nu A_\mu$ is the field strength and $\tilde{D}_\mu=\partial_\mu+ig \Aextdual+ig \tilde{A}_\mu$ is the dual covariant derivative. The indices $\mu$ and $\nu$ run over 1,2,3,4 and we keep all indices down for tensors in Euclidean signature.

Using the electromagnetic duality symmetry in the form $F_{\mu \nu}F_{\mu \nu} = \ti{F}_{\mu \nu}\ti{F}_{\mu \nu}$, we may dualise the photon kinetic term, writing the whole Lagrangian in terms of $\tilde{A}_\mu$. At this point, given the gauge field is integrated over, the tilde is merely notational and one may drop it entirely, resulting in the Lagrangian for scalar quantum electrodynamics (SQED) at strong coupling, except with external field $\Aextdual$ rather than $\Aext$. The result is that we calculate the Schwinger pair production of (strongly coupled) electrically charged particles from a time-dependent external electric field, but we refer to their charge as the magnetic charge $g$ and to the external field as the magnetic field $B$. The duality transformation in this case is simple because we do not treat the electrically charged particles of the heavy ions as dynamical --- their interactions with the magnetic monopoles are assumed to be entirely through $\Aextdual$.

We will assume in the following that the scalar self-coupling, $\lambda$, is sufficiently small that we may ignore it, at least in the range of energies considered. Of course photon loops will generate this term. However, the term is a point-like interaction between scalar loops (given no external legs) and, in the dilute instanton approximation that we will make, such loops are subdominant and are neglected. Note that for spin 1/2 elementary monopoles no such term would arise, the Euclidean Lagrangian being,
\begin{equation}
 \mathcal{L}_{s=1/2}:=\frac{1}{4}F_{\mu \nu}F_{\mu \nu} + \bar{\psi}(\slashed{D} + m)\psi, \label{eq:qedlagrangian}
\end{equation}
where the Feynman slash here denotes contraction with the 4D Euclidean gamma matrices (see Ref.~\cite{laine2016basics} for a definition). The dualisation of this Lagrangian is exactly as for the spin 0 case, resulting in quantum electrodynamics (QED) at strong coupling and with external field $\Aextdual$.

By purely formal manipulations, the partition functions for QED and SQED can be re-expressed exactly as path integrals over interacting worldlines \cite{feynman1951operator,affleck1981pair}. This representation is valid to all orders in $g$. For SQED it reads
\begin{widetext}
\begin{equation}
W_E = -\log \Bigg[1+
\sum_{n=1}^\infty\frac{1}{n!}\prod\limits_{a=1}^{n}\left( \int_0^\infty \frac{\mr{d}s_a}{s_a}\int\mathcal{D}x^a_\mu \ \mr{e}^{ -  S[x^a;s_a;\Aextdual]} \ \mr{e}^{ g^2 \sum_{b<a} \oint\oint \mr{d}x^a_\mu \mr{d}x^b_\nu G_{\mu\nu}(x^a,x^b) }\right)\Bigg]. \label{eq:W_worldline}
\end{equation}
\end{widetext}
The functional integrals over the $x_\mu^a:=x_\mu^a(\tau)$ are over closed worldlines in 4D Euclidean space. The $s_j$ are often referred to as Schwinger parameters and $G_{\mu\nu}$ is the free photon propagator. The Euclidean worldline action is given by,
\begin{align}
 S[x;s;\Aextdual] &= \frac{m^2s}{2} + \frac{1}{2s}\int^1_0\mr{d}\tau\dot{x}_\mu\dot{x}_\mu \nonumber \\
 &- ig\int_0^1 \Aextdual\dot{x}_\mu du \nonumber \\
 + \frac{g^2}{8\pi^2}&\int_0^1\mr{d}\tau\int_0^1\mr{d}\tau' \frac{\dot{x}_\mu(\tau) \dot{x}_\mu(\tau')}{|x(\tau) - x(\tau')|^2}, \label{eq:action}
\end{align} 
where in the last term we have inserted an explicit expression for $G_{\mu\nu}$ in a generic $R_\xi$ gauge and noted that the gauge dependent terms vanish for closed worldlines.

The last term in Eq.~\eqref{eq:action} is a double integral over the worldline, weighted with the photon propagator. It accounts for the interactions between different points on the worldline. For coincident points there is a UV divergence \cite{dotsenko1979renormalizability,polyakov1980gauge,brandt1981renormalization,polyakov1987gauge}, proportional to the length of the worldline and hence a power-like divergence. It can be identified with the usual UV-divergent contribution to the charged particle self-energy, which is removed by adding a mass counterterm. For worldlines without self-intersections or kinks, this is the only divergence of this term. Various regularisation schemes exist, and, just as in the field representation, the power-like divergence is absent in dimensionless regularisation schemes.

For spin 1/2 monopoles, Eqs.~\eqref{eq:W_worldline} and \eqref{eq:action} are modified by the addition of spin-dependent terms in the action \cite{feynman1951operator,polyakov1987gauge,Corradini:2015tik}. However, in the presence of weak external fields these terms are subdominant in powers of the weak field relative to the spin 0 part. For weak, constant, external fields, they contribute to the semiclassical prefactor, simply resulting in an overall factor of the number of degrees of freedom of the final state \cite{affleck1981pair,Gould:2018ovk}. For spacetime-dependent fields, the spin-dependent corrections could be more complicated than this and are worth understanding, but they are nevertheless subdominant so we do not treat them further here.

For sufficiently weak external fields, the probability of pair production is exponentially small. In this case, one finds that the pair production is well described by the leading nonzero term in a Virial (or cluster) expansion of Eq.~\eqref{eq:W_worldline}. Higher order terms are exponentially suppressed relative to this leading term. In the leading term, the imaginary part of the integral is dominated by a saddlepoint of the action, a worldline instanton $x_\mu^{\mr{inst}}$. The action 
 \begin{equation}
S_{\rm inst}
=S[x_\mu^{\mr{inst}};s^{\mr{inst}};\Aextdual]
 \end{equation}
of this worldline instanton gives the exponential suppression of the probability,
 \begin{equation}
{\rm Im}\,W_E\sim \mr{e}^{-S_{\rm inst}}
.
\end{equation}
 Integrating over quadratic fluctuations about the instanton gives the prefactor of the exponent, to leading order.

The combination of the semiclassical approximation and the Virial expansion are together referred to as the dilute instanton approximation, which we will make in what follows. The validity of this approximation relies on the dominance of a single worldline instanton in the integral over all worldlines. Thus we require the usual condition for semiclassicality,
\begin{equation}
    S[x_\mu^{\mr{inst}};s^{\mr{inst}};\Aextdual] \gg 1\, \label{eq:semiclassical_condition} 
\end{equation}
where $x_\mu^{\mr{inst}}$ and $s^{\mr{inst}}$ are the saddlepoint values, i.e. the worldline instanton. Further, we require that all scales of the worldline instanton are large compared with the scale on which small, virtual monopole-antimonopole pairs become important,
\begin{equation}
    \mr{Min}\left[R_C(x_\mu^{\mr{inst}}(\tau))\right] \gg \frac{\rcl}{2}, \label{eq:curvature_condition}
\end{equation}
where $R_C(x_\mu^{\mr{inst}}(\tau))$ is the radius of curvature (the inverse of the Gaussian curvature) of the worldline instanton at a point $\tau$. The size of virtual monopole-antimonopole pairs, $r$, be estimated by equating the rest mass of a monopole-antimonopole pair, $2m$, to their Coulomb attraction, $g^2/(4\pi r)$, resulting in $r=\rcl/2$. Equation \eqref{eq:curvature_condition} is important to ensure that the effects of virtual monopole-antimonopole pairs can be factored out of the instanton calculation, affecting only the running of couplings \cite{Gould:2018ovk}. Note that when this condition is satisfied, the scales of the worldline instanton are also much larger than the size of a dressed elementary monopole. We make no approximations with regard to the coupling, $g$.

So far in this section we have only discussed elementary monopoles. Solitonic monopoles are bound states of elementary fields and do not have their own local field operator appearing in the Lagrangian. Their size is generically of the same order as the classical monopole radius \cite{Coleman:1982cx}. Hence, when Eq.~\eqref{eq:curvature_condition} is met, solitonic monopoles are much smaller than the minimum radius of curvature of the worldline instanton. In this case solitonic monopoles can be described by an effective field theory identical to that of elementary monopoles~\cite{bardakci1978local,manton1978effective,affleck1981monopole,gould2017thermal}, meaning that our calculations are also applicable to them. This is because only the photon and graviton are massless and hence at long distances all monopoles with the same mass, spin and charge look the same.

\begin{figure*}
 \centering
  \includegraphics[width=1.0\textwidth]{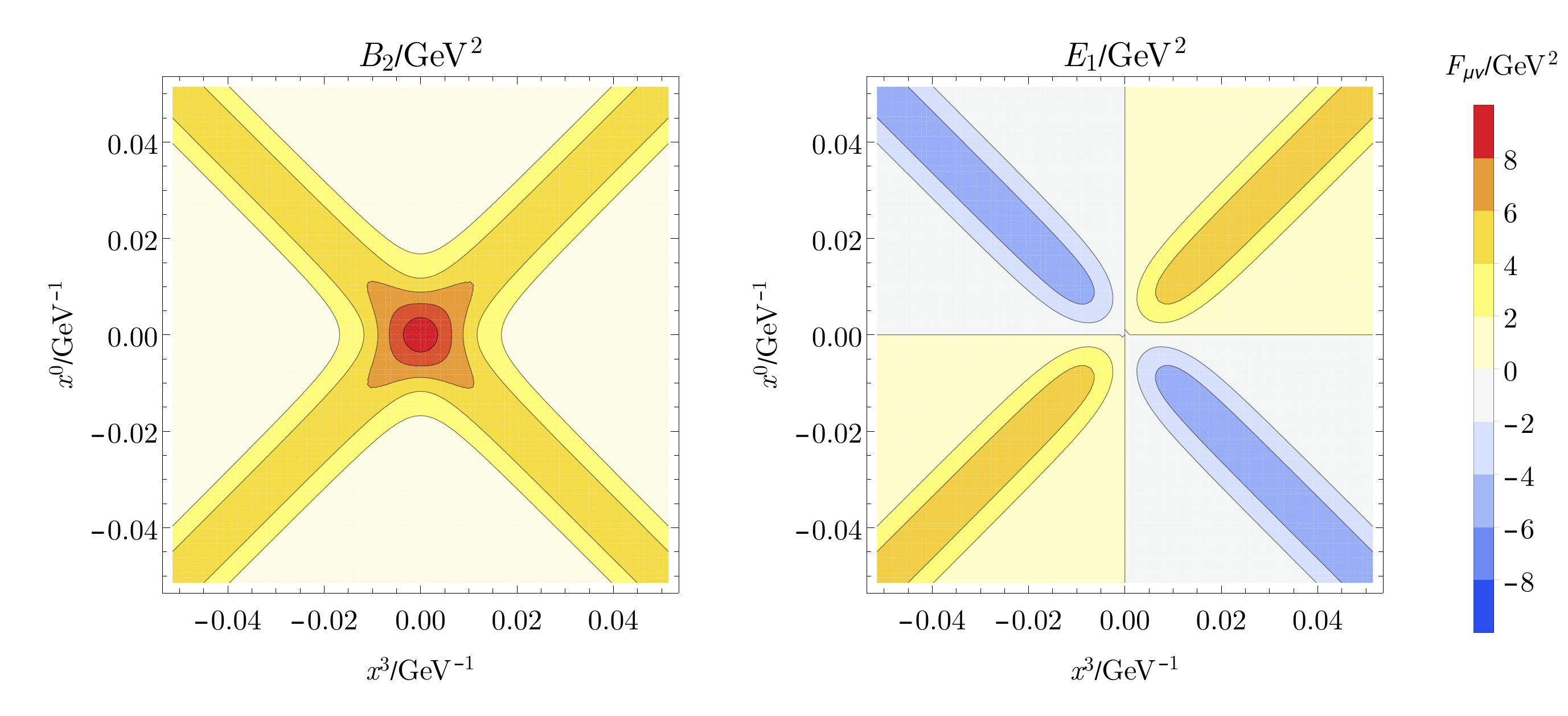}
  \caption{Magnetic field component, $B_2$  and electric field component, $E_1$, near origin of coordinate system for a collision of two lead ions with centre of mass energy per nucleon equal to 5.02TeV and impact parameter $2R$. Note that $x^1=x^2=0$ here.}
  \label{fig:fields}
\end{figure*}

\section{Electromagnetic fields}
\label{sec:em_fields}

With current and future magnetic monopole searches in mind, we will consider electromagnetic fields in heavy-ion collisions at ultrarelativistic energies. ALICE, ATLAS, CMS, LHCb, and MoEDAL may all be able to detect magnetic monopoles produced in heavy-ion collisions at the LHC, as their experimental signatures are extremely distinctive: they are highly ionising and follow parabolic tracks in uniform magnetic fields. In particular, the trapping detectors of MoEDAL are ideally suited for monopole detection because they have no background noise \cite{Acharya:2017cio}. For the most recent (2018) lead-ion collisions at the LHC, the centre of mass energy per nucleon, $\sqrt{s_{NN}}$, was equal to 5.02TeV, amounting to a very large Lorentz factor,
\begin{equation}
\gamma_{LHC}\approx \frac{\sqrt{s_{NN}}}{2m_p} \approx 2675.
\label{equ:gammaLHC}
\end{equation}
In the following, we will therefore assume that $\gamma \gg 1$.

The electromagnetic fields in ultrarelativistic heavy-ion collisions have been studied by many authors and are reviewed in Ref.~\cite{huang2015electromagnetic}. The electric fields of the ions are length contracted, being localised into an angular region of size $O(1/\gamma)$ about the perpendicular to the direction of motion \cite{landau1971classical}. A magnetic field is induced perpendicular to both the electric field and the direction of motion. In the ultrarelativistic limit, the magnetic field is of the same magnitude as the electric field. The strength of the electromagnetic fields are enhanced by the use of ions with large atomic numbers, $Z\gg 1$.

To describe the collision, we use lab coordinates with spacetime origin at the point of closest approach of the two ions. The beam axis points along the $x^3$ direction and the impact parameter points along the $x^1$ axis. Peripheral collisions (or even near-misses) lead to the largest magnetic fields. In this case there are only $O(1)$ participant nucleons in the collision and the spectator nucleons dominate the field up to $O(Z^{-1})$ corrections, which we drop \cite{kharzeev2007effects}. We do not include the conductivity of the ions as this does not significantly affect the fields at very early times when the magnetic field is greatest \cite{gursoy2014magnetohydrodynamics}. We also neglect quantum corrections to the heavy-ion generated fields, as these are expected to be small \cite{deng2012event}.

In the calculation of the electromagnetic field due to the ions, we adopt a mean-field approximation, treating the nuclei as a classical Woods-Saxon charge distribution,
\begin{equation}
\rho_\mr{WS}(r,R,a) = \frac{A}{1+\mr{e}^{(r-R)/a}}, \label{eq:saxon_woods}
\end{equation}
where $r$ is the position from the centre of the nucleus, the parameters $R$ and $a$ are taken from experiment and $A$ is merely a normalisation. For numerical evaluations, we adopt the values $R=6.62 \pm 0.06$fm and $a=0.546 \pm 0.010$fm for lead-ions, based on data from low-energy electron-nucleus scattering experiments \cite{DeJager:1987qc,alice2014centrality}.

The spectator nuclei move on inertial trajectories and hence, to calculate their electromagnetic fields, one need only boost the Coulomb field (or use the Li\'{e}nard-Wiechert potentials) and integrate over Eq.~\eqref{eq:saxon_woods}. At the spacetime origin of the coordinate system the electric field cancels, by symmetry, while the magnetic fields of the two nuclei double up. At this point the magnetic field points in the $x^2$ direction. Its magnitude here is the global maximum of all of the components of the electromagnetic field and is proportional to $\gamma$. It is the field in the neighbourhood of this point that is most likely to produce magnetic monopoles, if such particles exist. As such, it is important to know how the magnetic field dies off away from this point, and how the other components of the electromagnetic field increase.

By scaling the integrals determining the electromagnetic field, one can show the following parametric relations
\begin{align}
\frac{\partial F_{\mu\nu}}{\partial x^1} &\sim \frac{\partial F_{\mu\nu}}{\partial x^2} \sim \frac{F_{\mu\nu}}{R}, \nonumber \\
\frac{\partial F_{\mu\nu}}{\partial x^3} &\sim \frac{\partial F_{\mu\nu}}{\partial x^0} \sim \frac{F_{\mu\nu}}{R/\gamma}, \label{eq:derivatives}
\end{align}
where $F_{\mu\nu}$ is any component of the electromagnetic field and we have taken $R \sim b \sim a$, in that all are proportional to zero powers of $\gamma$. Thus for very large $\gamma$ the electromagnetic fields are localised to a region of size $O(R/\gamma)$ in the $x^3$ and $x^0$ directions and of size $O(R)$ in the $x^1$ and $x^2$ directions. As we will see, the pair production process is localised to within a region of size $O(R/\gamma)$, hence we can drop dependence on the $x^1$ and $x^2$ directions and focus on the $x^3$ and $x^0$ dependence.

\begin{figure}
 \centering
  \includegraphics[width=0.45\textwidth]{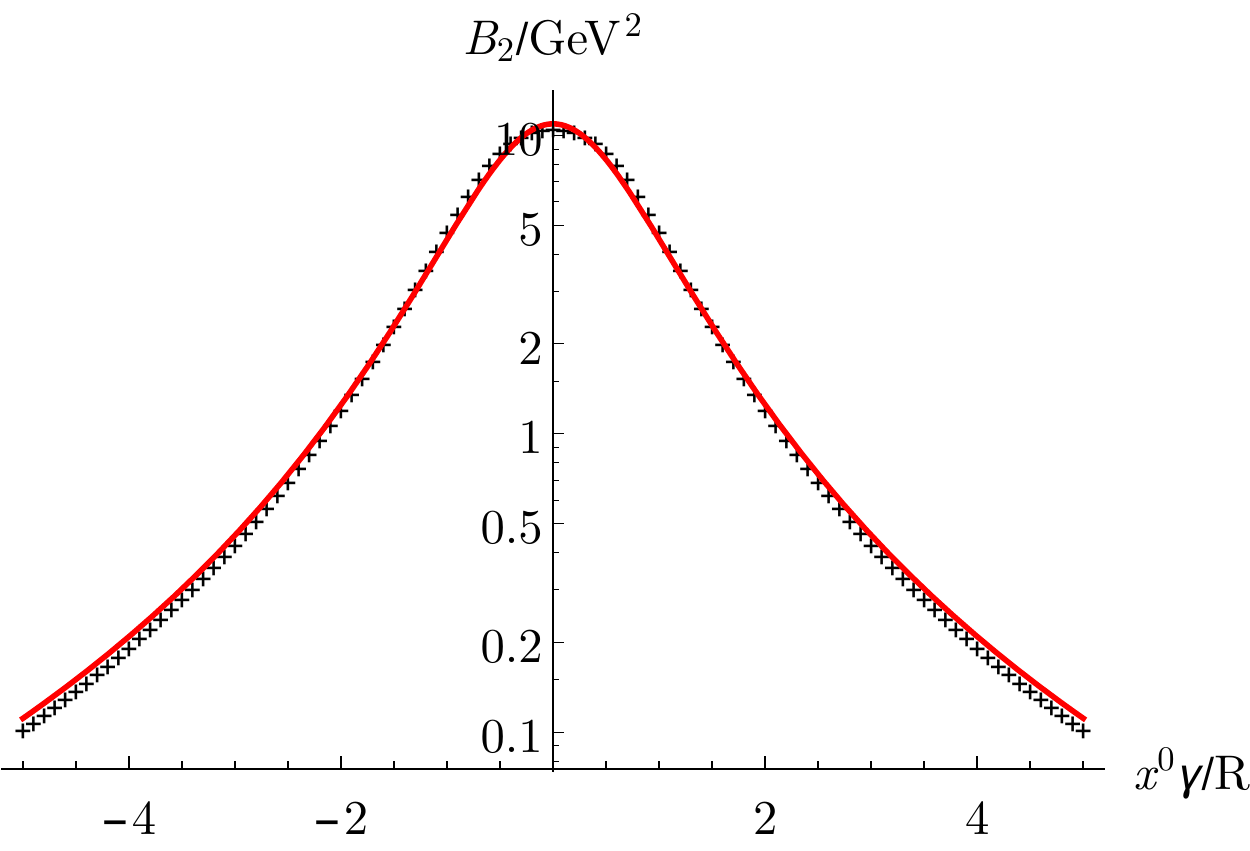}
  \caption{Plot of the magnetic field, $B_2$, at the spatial origin of the coordinates for a collision of two lead ions with centre of mass energy per nucleon equal to 5.02TeV and impact parameter $2R$. Note that $x^1=x^2=x^3=0$ here. Our fit, Eq.~\eqref{eq:fit_functions}, is shown as a continuous red line alongside the results of performing the numerical integrals, as black crosses.}
  \label{fig:Bfit}
\end{figure}

\begin{figure}
 \centering
  \includegraphics[width=0.45\textwidth]{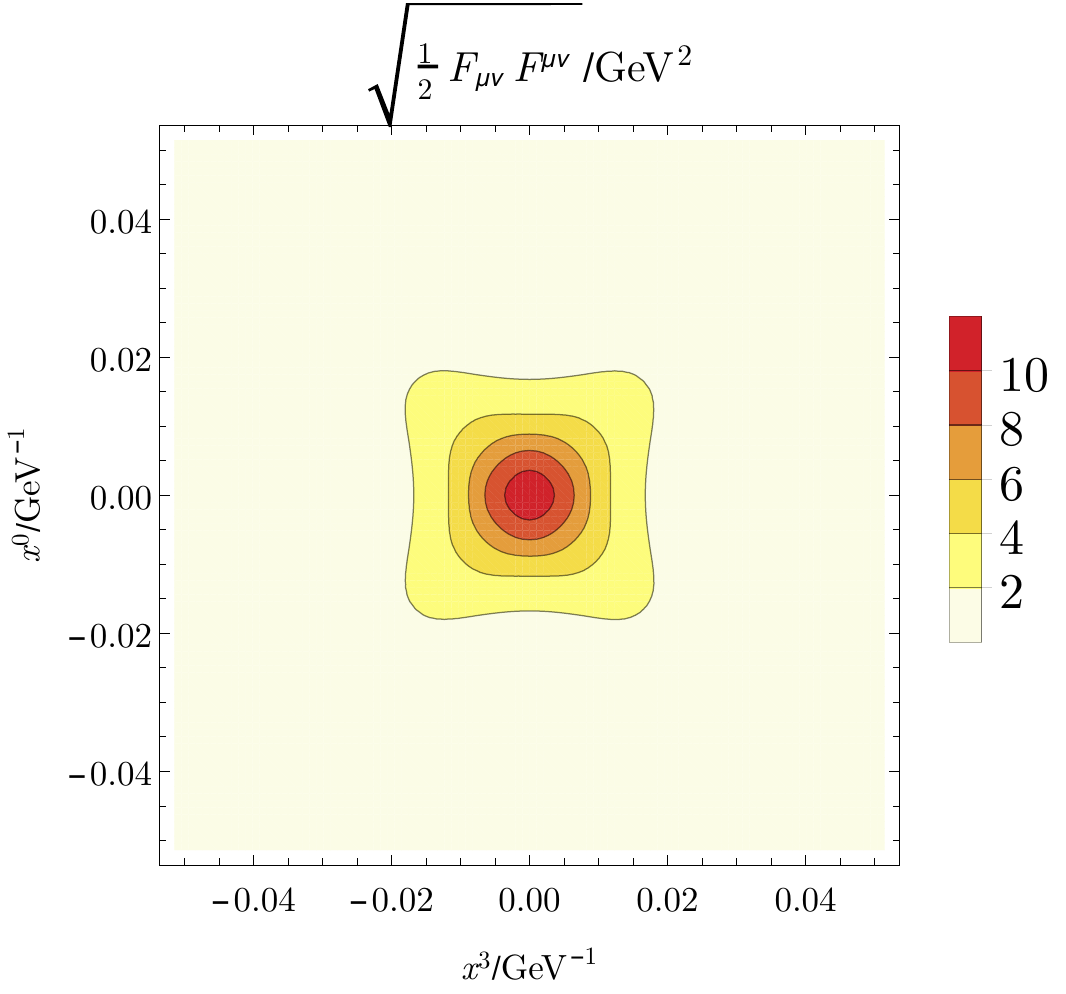}
  \caption{Plot of the nonzero scalar invariant of the electromagnetic field, $\tfrac{1}{2}F_{\mu\nu}F^{\mu\nu}=\mb{B}^2-\mb{E}^2$. In this plane, the other scalar invariant, $\tfrac{1}{4}F_{\mu\nu}\ti{F}^{\mu\nu}=\mb{E}\cdot \mb{B}$, is zero and away from this plane it is suppressed relatively by $\gamma$.}
  \label{fig:invariants}
\end{figure}

In this case only two components of the electromagnetic field are nonzero, $B_2$ and $E_1$. The electromagnetic dual of this field configuration is given by $\ti{E}_2 = B_2$ and $\ti{B}_1 = -E_1$. The results of performing the integrals of the Li\'{e}nard-Wiechert potentials over the Woods-Saxon distributions are shown in Fig.~\ref{fig:fields}. The corresponding scalar invariants are given in Fig.~\ref{fig:invariants}. Inspired by the field configurations for point-like charges, we find that the results can be well approximated by
\begin{widetext}
\begin{align}
B_2 &= \frac{B}{2}\left(\frac{1}{\left(1+\omega^2(x^0-x^3/v)^2\right)^{3/2}}+\frac{1}{\left(1+\omega^2(x^0+x^3/v)^2\right)^{3/2}}\right), \nonumber \\
E_1 &= \frac{B}{2}\left(\frac{1}{\left(1+\omega^2(x^0-x^3/v)^2\right)^{3/2}}-\frac{1}{\left(1+\omega^2(x^0+x^3/v)^2\right)^{3/2}}\right), \label{eq:fit_functions}
\end{align}
\end{widetext}
where $B$ is the value of the magnetic field at the spacetime origin, $v\approx 1$ is the ion speed and $\omega$ is a fit parameter. Both depend on the particular heavy-ion collision considered through $b$ and $\gamma$ (or $\sqrt{s}$). In Fig.~\ref{fig:Bfit} we show the magnetic field along with our fit at $x^3=0$. Relative deviations from our fit are only a few percent, so we do not complicate our fit function to account for them.

The largest cross section for pair production will occur for the largest values of the magnetic field. For $b\lesssim 2R$, $B$ increases linearly with $b$, reaching a maximum at $\bmax\approx 1.94R$ before decreasing again. The value of $\bmax$ can be shown to be independent of $\gamma$. About this maximum, we find
\begin{align}
B(b,\gamma)&= B(\bmax,\gamma)\nonumber \\
&\ \left(1 -\frac{1}{2}\frac{c_{B2}}{R^2}(b-\bmax)^2+O(b-\bmax)^3\right),
\end{align}
where the numerical coefficient $c_{B2}\approx 1.37$ is found by a quadratic fit to the numerical data and, like $\bmax$, is independent of $\gamma$.

For fixed $b$, the magnetic field is a linearly increasing function of $\gamma$. For $b=\bmax$, we find
\begin{equation}
 B(\bmax,\gamma) \approx c_B \frac{Ze v \gamma}{2\pi R^2},\label{eq:Bmaxgamma}
\end{equation}
where we have written the result in terms of that for point-like ions, and the numerical coefficient $c_B\approx 0.78$ is independent of $\gamma$.

The second parameter of the fit, $\omega$, is of order $\gamma/R$, as is clear from Eqs.~\eqref{eq:derivatives}. We find
\begin{equation}
 \omega(\bmax,\gamma) \approx c_\omega \frac{v \gamma}{R},\label{eq:omegamaxgamma}
\end{equation}
where the numerical coefficient $c_\omega\approx 0.92$ is independent of $\gamma$. For $b\lesssim \bmax$ we find that $\omega$ is approximately independent of $b$, whereas for $b\gtrsim\bmax$ it decreases approximately linearly,
\begin{align}
\omega(b,\gamma)&\approx \omega(\bmax,\gamma)  \nonumber \\
&\ \left(1-\frac{c_{\omega 1}}{R}\theta(b-\bmax)(b - \bmax)\right),
\end{align}
where $c_{\omega 1}\approx 0.25$ and is independent of $\gamma$ for $\gamma\gtrsim 5$. Of course the transition is not as sharp as the step function suggests, but is smoothed over a region of size $a$ (see Eq.~\eqref{eq:saxon_woods}). That $\omega$ is smaller for $b\gtrsim \bmax$ than for $b\lesssim\bmax$ will lead to a reduction of the production cross section for near misses with respect to peripheral collisions. 

In our calculations of the fields, we have not included event-by-event fluctuations in nucleon positions, which cause deviations from our mean-field results \cite{deng2012event}. However, inclusion of these effects does not change the scaling relations outlined above: the field variation in the transverse direction varies over a larger distance (by a factor of \(\sim \gamma\)) than the longitudinal fluctuations, and the approximate analytic expression for the time dependence still holds. Furthermore, at the spacetime origin, the longitudinal component of the magnetic field, and the components of the electric field, remain at least an of magnitude smaller than the transverse magnetic field.

\section{The worldline instanton} \label{sec:worldlineInstanton}
In this section we determine the worldline instanton for the field in a high-energy heavy-ion collision, and calculate the corresponding exponential dependence of the pair production probability. This is given by
\begin{equation} \label{eq:probabilityExpression}
	P \approx D \mr{e}^{-S},
\end{equation}
where \(S\) is the classical action \eqref{eq:action} evaluated at its saddlepoint, and \(D\) is the semiclassical prefactor, given by a functional determinant.

The fields of interest are those given in \eqref{eq:fit_functions}. To find instantons we perform a Wick rotation \(x^0 \to i x_4\), yielding the Euclidean fields
\begin{widetext}
\begin{align}
B_2^E &= -\frac{i B}{2}\left(\frac{1}{\left(1+\omega^2(i x_4-x_3/v)^2\right)^{3/2}}+\frac{1}{\left(1+\omega^2(i x_4+x_3/v)^2\right)^{3/2}}\right), \nonumber \\
E_1^E &= \frac{B}{2}\left(\frac{1}{\left(1+\omega^2(i x_4-x_3/v)^2\right)^{3/2}}-\frac{1}{\left(1+\omega^2(i x_4+x_3/v)^2\right)^{3/2}}\right).
\end{align}
\end{widetext}
The extra factor of \(-i\) in the magnetic field is a conventional choice accounting for the derivative with respect to imaginary time in the definition of the (dual) field tensor \cite{affleck1981pair}. 
For these specific fields, it makes both $E^E$ and $B^E$ purely imaginary. Therefore the Euclidean worldline instanton equations are purely real,
\begin{equation} \label{eq:euclideanEoms}
    \ddot{x}_\mu = -i g s \tilde{F}_{\mu \nu}^E \dot{x}_\nu,
\end{equation}
where
\begin{equation}
 \tilde{F}_{\mu \nu}^E =
 \left( \begin{array}{cccc}
0 & 0 & 0 & 0  \\
0 & 0 & -E_1^E & B_2^E  \\
0 & E_1^E & 0 & 0  \\
0 & -B_2^E & 0 & 0  
\end{array} \right)
\end{equation}
and the indices $\mu$ and $\nu$ run over 1,2,3,4, with the 4 component last.

Instanton solutions take the form of closed solutions to the Euclidean equations of motion. From the symmetry of the field it is clear that such a solution exists in the plane \(x_1 = x_3 = 0\), where the fields reduce to
\begin{align} \label{eq:1dBField}
	B_2^E(x_{1,3} = 0) &= \frac{-i B}{\left(1 - (\omega x_4)^2\right)^{3/2}}, \\
	E_1^E(x_{1,3} = 0) &= 0. \nonumber
\end{align}
The instanton equations then reduce to those in a purely time-dependent magnetic field. This feature of instantons in fields where the spatial variation is perpendicular to the direction of the field has been noted previously in Ref.~\cite{torgrimsson2016schwinger}. As the exponential dependence of the pair production probability is determined completely by the action of the worldline instanton, the effects of the inhomogeneity in the transverse spatial directions will only contribute to the production probability at the level of the prefactor. This considerably simplifies the problem of computing the pair production probability: Schwinger production in fields that vary along a single spacetime dimension have been widely studied \cite{brezin1970pair,Nikishov:1970br,Popov:1971ff,popov1972pair,Marinov1977electron,dunne2005worldline,dunne2006worldline,kim2006schwinger}.

Henceforth for notational convenience all fields will be implicitly Euclidean unless otherwise indicated.

Following Ref.~\cite{affleck1981pair} we treat the worldline self-interaction term separately, writing the action \eqref{eq:action} as
\begin{equation}
    S[x_\mu, s] = S_0[x_\mu, s] + \Delta S[x_\mu]
\end{equation}
where
\begin{align}
    S_0[x_\mu, s] &:= \frac{m^2s}{2} + \frac{1}{2s}\int^1_0\mr{d}\tau\dot{x}_\mu\dot{x}_\mu \nonumber \\
    &\qquad\qquad\qquad - ig\int_0^1 \mr{d} \tau \, \Aextdual\dot{x}_\mu, \\
    \Delta S[x_\mu] &:= \frac{g^2}{8\pi^2}\int_0^1\mr{d}\tau\int_0^1\mr{d}\tau' \frac{\dot{x}_\mu(\tau) \dot{x}_\mu(\tau')}{|x(\tau) - x(\tau')|^2}.\label{eq:action_0} 
\end{align}
In Sections \ref{sec:instanton_free} and \ref{sec:corrections} we assume that \(|\Delta S| \ll |S_0|\) when evaluated at the saddle point. Note that this is not a perturbative expansion in \(g\); the precise conditions for this relation to hold will be examined at the end of Section \ref{sec:corrections}. In Section \ref{sec:numerics} we perform a full calculation treating \(\Delta S\) to all orders, numerically.

\subsection{Worldline instanton without self-interactions}\label{sec:instanton_free}
It is convenient to choose a gauge such that the dual electromagnetic potential is
\begin{equation}
    \Aextdual = \frac{i B x_4}{\sqrt{1 - (\omega x_4)^2}} \delta_{\mu 2}.
\end{equation}
Note that Eq.~\eqref{eq:action_0} is gauge invariant, due to the worldline being closed, so we are free to choose a gauge. Ignoring the self-interaction term the worldline instanton stationarises
\begin{align} \label{eq:noninteractingAction}
    S_0[x_\mu, s] &= \frac{m^2s}{2} + \frac{1}{2s}\int^1_0\mr{d}\tau\dot{x}_\mu\dot{x}_\mu \nonumber \\
    &\qquad + gB\int_0^1 \mr{d} \tau \frac{\dot{x}_2 x_4}{\sqrt{1 - (\omega x_4)^2}}.
\end{align}
Worldline actions of this form have been extensively studied by Dunne et.~al \cite{dunne2005worldline,dunne2006worldline}. In order to follow the general prescription outlined in Ref.~\cite{dunne2006worldline} (motivated by the work of Keldysh on ionisation in inhomogeneous fields \cite{keldysh1965ionization}) we define the the dimensionless \emph{Keldysh parameter} 
\begin{equation}
    \kp := \frac{m \omega}{g B}.
    \label{equ:xidef}
\end{equation}
We choose to use \(\kp\) instead of the more conventional \(\gamma\) in order to avoid confusion with the Lorentz factor. The physical interpretation of \(\kp\) when considering monopole production in heavy-ion collisions is discussed in Section \ref{sec:consequences}. 

\begin{figure}
 \centering
  \includegraphics[width=0.45\textwidth]{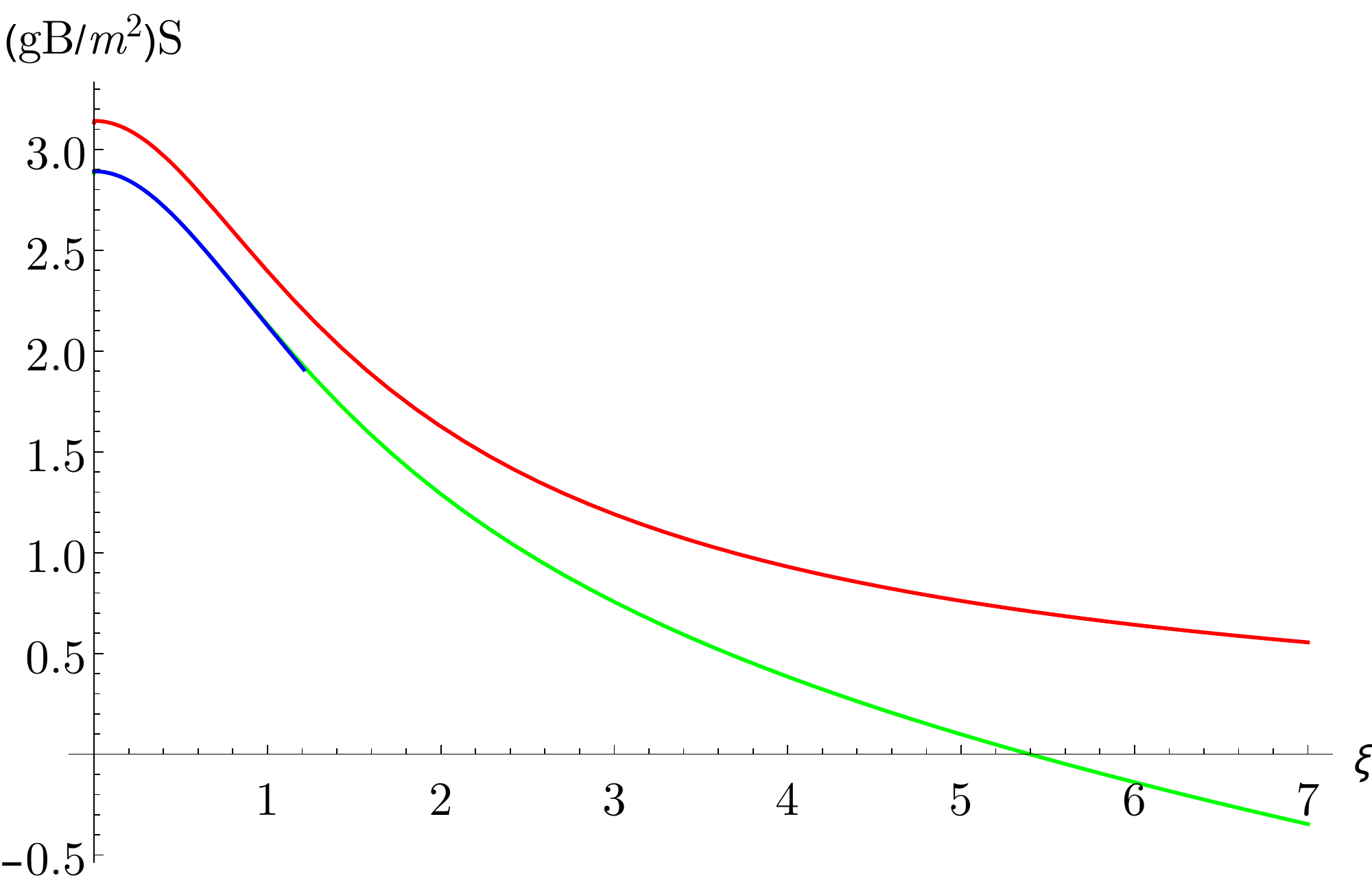}
  \caption{Plot of the worldline instanton action as a function of the Keldysh parameter in various levels of approximation. The red line is the result without self-interactions, Eq.~\eqref{eq:action_dunne}, the green line includes the leading correction from self-interactions, Eq.~\eqref{eq:leading_correction}, and the blue line gives the numerical, all-orders results of Section \ref{sec:numerics}, for $g^3B/m^2=1$.
  }
  \label{fig:action_slice}
\end{figure}

It was shown in Refs.~\cite{dunne2005worldline,dunne2006worldline} (in the context of electron-positron pair production) that, at the saddle point, the non-self-interacting action \eqref{eq:noninteractingAction} evaluates to
\begin{align}
    S_0[x^{(0)}_\mu] &= \frac{2 m^2}{g B} \int_{-1}^1 \mr{d} y \frac{\sqrt{1 - y^2}}{(1 + \kp^2 y^2)^\frac{3}{2}} \nonumber \\
    & = \frac{4 m^2}{g B \kp^2}[\mb{E}(-\kp^2) - \mb{K}(-\kp^2)],\label{eq:action_dunne}
\end{align}
where \(\mb{E}\) and \(\mb{K}\) are elliptic integrals, and \(x^{(0)}_\mu\) denotes the worldline instanton for the non-self-interacting action (detailed below). This result is shown as the red curve in Fig.~\ref{fig:action_slice}.

As \(\kp \to 0\),
\begin{equation}
    S_0[x^{(0)}_\mu] \to \frac{\pi m^2}{g B};
\end{equation}
the constant-field result is obtained. For a rapidly varying field (\(\kp \gg 1\)),
\begin{equation}
    S_0[x^{(0)}_\mu] \to \frac{4 m^2}{g B \kp} = \frac{4 m}{\omega}.
\end{equation}
The functional form of the pair production probability --- notably the mass dependence --- changes in the limit of strongly time-dependent fields. This has important implications for the production of high-mass monopoles in heavy-ion collisions, discussed in Section \ref{sec:consequences}.

In Refs.~\cite{dunne2005worldline,dunne2006worldline} the fluctuation prefactor \(D\) (see Eq.~\eqref{eq:probabilityExpression})  for fields of the form \eqref{eq:1dBField} is also calculated, and is given approximately by
\begin{align}
	D &\approx (2s+1) V_3 \frac{\sqrt{2 \pi}(g B)^{3/2}}{32 \pi^2} \nonumber \\
	&\ \frac{(1 + \kp^2)^{3/4}}{\mb{E}(-\kp^2) \sqrt{(1 + \kp^2) \mb{K}(-\kp^2) - (1 - \kp^2) \mb{E}(-\kp^2)}},
\end{align}
where $s$ is the monopole spin and \(V_3\) is the spatial volume factor. However, for our case the $x_3$ dependence of the field will modify the prefactor to leading order in $\gamma$. This is because the prefactor involves the determinant of fluctuations about the instanton, and fluctuations in the $x_3$ direction will feel this dependence. A full calculation of the prefactor for our fields should be possible and is planned for further work.

However, for the purpose of obtaining order-of-magnitude estimates, we note that the \(\kp\)-dependent part of the prefactor is equal to \(1/\kp\) to within an \(O(1)\) factor for all \(\kp\): the prefactor is of the same order as that in the locally constant field approximation (LCFA) regardless of the magnitude of the Keldysh parameter. Noting this, we propose using the LCFA to approximate the prefactor also in the spatial directions (see Appendix \ref{appendix:lcfa} for details). In this approximation the curvature of the field at its maximum determines the prefactor. Denoting the much slower decay rate of the field in the $x_1$ and $x_2$ directions as $\Omega\ll \omega$, we therefore expect
\begin{equation} \label{eq:approxPrefactor}
	D \sim \frac{(2s+1)(g B)^4}{18 \pi ^3 m^4 \omega^2 \Omega^2},
\end{equation}
to provide a reasonable estimate of the prefactor, up to an $O(1)$ multiplicative factor.

\begin{figure}
 \centering
  \includegraphics[width=0.42\textwidth]{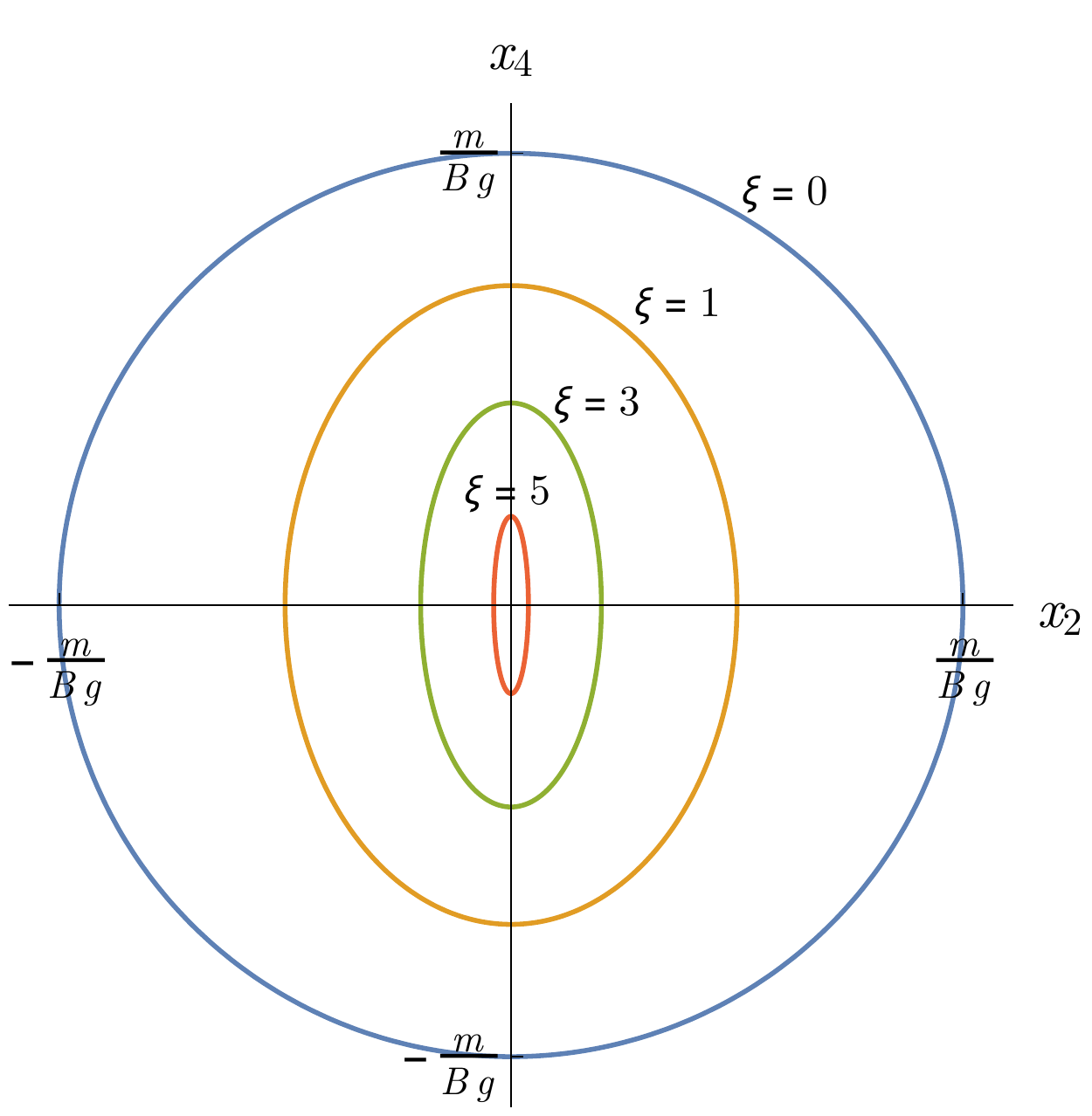}
  \caption{Elliptical worldlines stationarising the non-self-interacting action \eqref{eq:noninteractingAction} for different values of the Keldysh parameter, \(\kp\).}
  \label{fig:worldlinePlots}
\end{figure}

The shape of the worldline can be determined using a method closely related to that used in Ref.~\cite{dunne2005worldline}. Contracting the Euclidean equations of motion \eqref{eq:euclideanEoms} with \(\dot{x}_\mu\) shows that \(\dot{x}_\mu \dot{x}_\mu\) is a constant of motion, and varying the action with respect to the Schwinger parameter \(s\) shows that its saddlepoint value satisfies
\begin{equation} \label{eq:constantS}
    s^2 = \dot{x}_\mu \dot{x}_\mu.
\end{equation}
Using the symmetry properties of the field, Eq.~\eqref{eq:euclideanEoms} simplifies significantly; the non-trivial relations remaining are
\begin{align}
    \ddot{x}_2 &= \frac{g B}{m}\frac{s \dot{x}_4}{[1 - (\kp x_4)^2]^{3/2}}, \label{eq:x2Eom} \\
    \ddot{x}_4 &= -\frac{g B}{m} \frac{s \dot{x}_2}{[1 - (\kp x_4)^2]^{3/2}}, \label{eq:x4Eom} \\
    s^2 &= (\dot{x}_2)^2 + (\dot{x}_4)^2. \label{eq:constantSCl}
\end{align}
Integrating Eq.~\eqref{eq:x2Eom} gives
\begin{equation} \label{eq:integratedX2Eom}
    \dot{x}_2 = \frac{g B}{m} \frac{s x_4}{\sqrt{1 - (\omega x_4)^2}},
\end{equation}
and combining this with Eq.~\eqref{eq:constantSCl} gives{}
\begin{equation} \label{eq:x4ODE}
    (\dot{x}_4)^2 = s^2 \left(1 - \frac{x_4}{\sqrt{1 - (\omega x_4)^2}}\right).
\end{equation}
This can be integrated directly to give an explicit proper-time parametrisation of \(x_4(\tau)\) and \(x_2(\tau)\) in terms of Jacobi elliptic functions. However, the shape of the worldline in the \(x_2\)-\(x_4\) plane can be seen more clearly from the implicit expression
\begin{equation}
    \left(\frac{\mr{d} x_4}{\mr{d} x^2} \right)^2 = \frac{s^2 - (\dot{x}_2)^2}{(\dot{x}_2)^2} = \frac{s^2}{(\dot{x}_2)^2} - 1.
\end{equation}
Substituting Eq.~\eqref{eq:integratedX2Eom} gives
\begin{equation} \label{eq:noninteractingInstanton}
    \left(\frac{\mr{d} x_4}{\mr{d} x_2} \right)^2 = \left(\frac{m}{g B}\right)^2 \frac{1}{(x_4)^2} - (\kp^2 + 1).
\end{equation}
This can be readily checked to describe an ellipse: comparison with standard expressions gives the semi-major axis aligned along \(x_4\):
\begin{equation}
    a_4 = \frac{m}{gB}\frac{1}{\sqrt{1 + \kp^2}},
\end{equation}
and the semi-minor axis aligned along \(x_2\):
\begin{equation}
    a_2 = \frac{m}{gB}\frac{1}{1 + \kp^2}.
\end{equation}
In corroboration with results from previous analyses \cite{dunne2005worldline, dunne2006worldline}, the time dependence of the magnetic field contracts the worldline instanton and increases its departure from the circular constant-field result. The time dependence of the field can be parametrised by the Keldysh parameter \(\kp\), and the constant-field result is obtained smoothly in the limit \(\kp \to 0\). Plots of the non-self-interacting worldline instanton for different values of the Keldysh parameter are shown in Figure \ref{fig:worldlinePlots}.

\subsection{Self-interactions to leading order}
\label{sec:corrections}
Section \ref{sec:instanton_free} was largely a reproduction of known results for Schwinger production in time-dependent fields. In this and the following section we extend the calculation to account for worldline self-interactions. In this section we start by considering just the leading order corrections from worldline self-interactions, which give a contribution to the action
\begin{equation} \label{eq:interactionTerm}
    \Delta S[x_\mu] = \frac{g^2}{8\pi^2}\int_0^1\mr{d}\tau\int_0^1\mr{d}\tau' \frac{\dot{x}_\mu(\tau) \dot{x}_\mu(\tau')}{|x(\tau) - x(\tau')|^2}.
\end{equation}
This self-interaction term was originally studied in a similar context in Ref.~\cite{affleck1981pair}, where they considered a constant external field. As the non-self-interacting worldline instanton \eqref{eq:noninteractingInstanton} stationarises Eq.~\eqref{eq:noninteractingAction}, the leading order correction can be computed by evaluating \(\Delta S\) over the elliptical worldline described by Eq.~\eqref{eq:noninteractingInstanton}.

The self-interaction term is independent of the choice of worldline parametrisation, so we may choose to parametrise the non-self-interacting worldline instanton \(x^{(0)}_\mu\) in terms of the cylindrical polar angle \(\theta = \tan^{-1}(x_4/x_2)\):
\begin{equation}
    x^{(0)}_\mu(\theta) = \frac{m}{gB(1 + \kp^2)}(0,\cos \theta,0,\sqrt{1 + \kp^2}\sin \theta).
\end{equation}
\begin{widetext}
With this parametrisation the leading order correction may be expressed by
\begin{equation}
    \Delta S[x^{(0)}_\mu] = \frac{g^2}{8 \pi^2} \int_0^{2 \pi} \mr{d} \theta \int_0^{2 \pi} \mr{d} \theta' \frac{\cos \theta \cos \theta' + (1 + \kp^2) \sin \theta \sin \theta'}{(1 + \kp^2) [\cos \theta - \cos \theta']^2 + [\sin \theta - \sin \theta']^2}.
\end{equation}
This integral may be expressed as a double contour integral in the complex plane by performing the substitutions \(z = \mr{e}^{i \theta}\), \(z' = \mr{e}^{i \theta'}\):
\begin{equation}
    \Delta S[x^{(0)}_\mu] = \frac{g^2}{8 \pi^2} \oint_{|z| = 1} \mr{d} z \oint_{|z'| = 1} \mr{d} z' \frac{(z^2 + 1)({z'}^2 + 1) - (1 + \kp^2)(z^2 - 1)({z'}^2 - 1)}{(z - z')^2[1 - zz' + (1 + \kp^2)(1 + zz')][1 + zz' - (1 + \kp^2)(1 - zz')]}.
\end{equation}
\end{widetext}
The integral can now be performed using the residue theorem. As the integrand is explicitly symmetric under \(z \leftrightarrow z'\) the order of integration is unimportant. The pole at \(z = z'\) corresponds to the expected divergence from coincident points \cite{dotsenko1979renormalizability,polyakov1980gauge,brandt1981renormalization}, which may be removed by adding a mass counterterm as previously discussed in Section \ref{sec:general_approach}. After subtracting this divergence, and noting that \(\kp > 0\) for all physical cases, we find
\begin{equation}
    \Delta S[x^{(0)}_\mu] = -\frac{g^2}{8}\left(\sqrt{1 + \kp^2} + \frac{1}{\sqrt{1 + \kp^2}}\right). \label{eq:leading_correction}
\end{equation}
This tends to the known result for the circular worldline
\cite{affleck1981pair}, in the constant-field limit \(\kp \to 0\),
\begin{equation}
\lim_{\xi\rightarrow 0} \Delta S[x^{(0)}_\mu]=-\frac{g^2}{4}.
\label{equ:circularDeltaS}
\end{equation}
We have also verified its agreement with a numerical evaluation of the integral with an explicit short-distance regularisation and counterterm following Ref.~\cite{polyakov1980gauge}. It qualitatively matches a numerical evaluation of the correction for fields with a similar time dependence presented in \cite{lan2018holographic}, universally enhancing production probability, with a stationary point at \(\kp = 0\) and linear \(\kp\) dependence in the large-\(\kp\) limit. As in the constant-field case, the leading order self-interaction term is scale-invariant; it is only a function of worldline shape.

The exponential dependence of the monopole pair production probability in a high-energy heavy-ion collision is thus, to first order in the worldline self-interaction,
\begin{align} \label{eq:fullProbability}
    \ln P &\sim -\frac{\pi m^2}{g B}\frac{4 [\mb{E}(-\kp^2) - \mb{K}(-\kp^2)]}{\pi \kp^2} \nonumber \\
    &\qquad +\frac{g^2}{8}\left(\sqrt{1 + \kp^2} + \frac{1}{\sqrt{1 + \kp^2}}\right).
\end{align}
This is shown as the green curve in Fig.~\ref{fig:action_slice}.

Examining the limits of this expression highlights the conditions under which the assumption \(|\Delta S| \ll |S_0|\) is valid: as \(\kp \to 0\) we retain the constant-field case, 
where the condition is
\begin{equation}
    \frac{g^3 B}{4\pi m^2} \ll 1.
\end{equation}
However, for strictly constant fields, all higher order corrections vanish due to symmetry~\cite{affleck1981pair}, and hence this condition is in fact not necessary. For \(\kp \gg 1\), the condition becomes
\begin{equation} \label{eq:highKpPertubativityCondition}
   \frac{g^3B\xi^2}{32m^2} = \frac{g \omega^2}{32 B} \ll 1.
\end{equation}
Note that both of these conditions may be achieved for any value of the monopole charge, \(g\); the application of perturbation theory in the self-interactions does not require weak coupling.
On the other hand, condition (\ref{eq:highKpPertubativityCondition}) always fails at high enough $\xi$, indicating that the leading-order self-interaction correction is then no longer sufficient.

\subsection{Self-interactions to all orders}\label{sec:numerics}

Going beyond treating the self-interactions perturbatively, in this section we present our calculation of the worldline instantons taking self-interactions into account to all orders. In this case the equations of motion are integro-differential, due to the nonlocal nature of the self-interactions. Due to the lack of symmetries, these equations are rather hard to solve and hence we resort to a numerical approach, following Ref.~\cite{gould2017thermal} (see also Ref.~\cite{Schneider:2018huk}). We discretise the worldline, approximating it by a finite but large number of points, $N\gg 1$. The equations of motion are then simply $N$ nonlinear algebraic equations which we solve iteratively, using the Newton-Raphson method.

The self-interaction is singular at short distances, and hence needs regularisation. We follow the approach of Ref.~\cite{polyakov1980gauge} and introduce an explicit cut-off scale, $a$. However, for numerical stability we modify the counterterm following Ref.~\cite{gould2017thermal} (see Appendix \ref{appendix:discretisation} for details). By solving the equations of motion for a range of cut-off scales, we can then extrapolate to the $a\to 0$ limit, which we do following Ref.~\cite{gould2017thermal}. The explicit discretisation of the action that we use is given in Appendix \ref{appendix:discretisation}.

The number $N$ must be chosen such that the distance between neighbouring points, $|dx^i|:=|x^{i+1}-x^i|$, is much smaller than the smallest scale in the problem, the cut-off, $a$. Note that for a continuous worldline, the global reparametrisation symmetry $\tau\to\tau+c$ means that $\dot{x}_\mu\dot{x}_\mu$ is constant. Thus, to leading order in $1/N$, $|dx^i|$ is independent of $i$ and hence equal to $L[x]/N$, where $L[x]$ is the length of the loop. Further, the cut-off $a$ must be chosen to be much smaller than any other scale in the problem. In summary we require
\begin{equation}
 \frac{L[x]}{N}\ll a \ll \mr{Min}\left[\kappa,R_C(x;i)\right], \label{eq:scale_hierarchy}
\end{equation}
where $R_C(x;i)$ is the radius of curvature of the worldline at the point $i$. We mostly used $N=2^{12}$ points to describe the worldlines, though we also compared this to other values of $N$ in checking the $N\to \infty$ behaviour.

\begin{figure}
 \centering
  \includegraphics[width=0.45\textwidth]{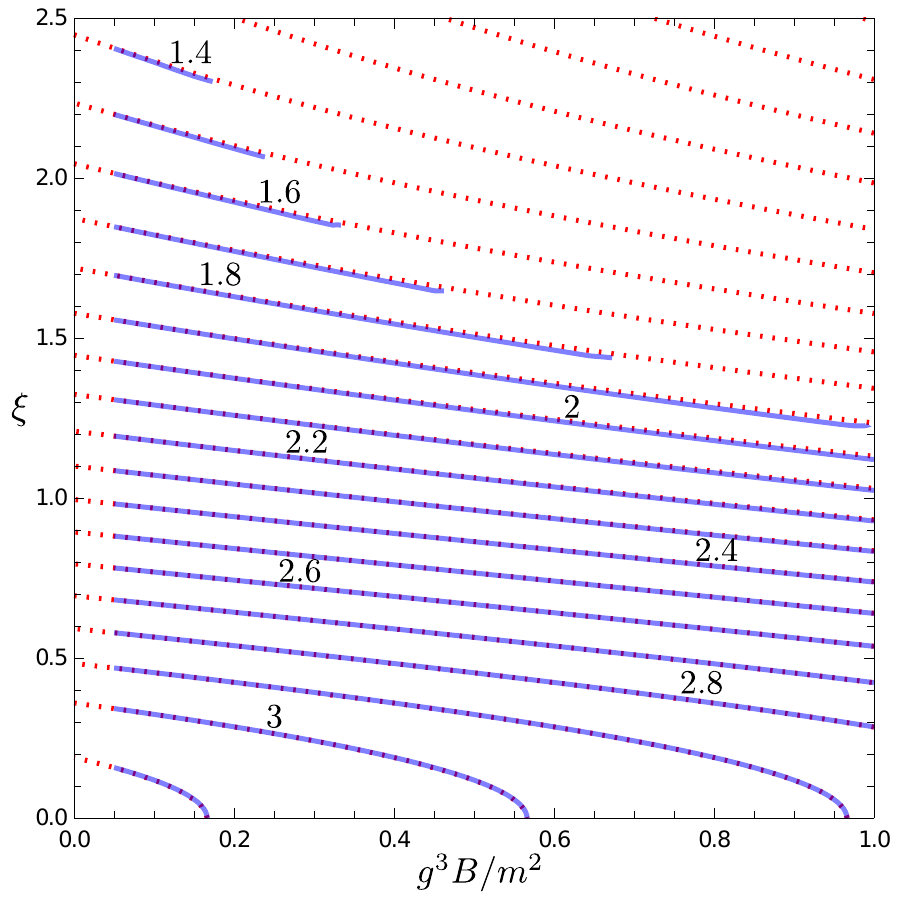}
  \caption{The worldline instanton action, $S$, scaled by $gB/m^2$. The contour plot shows the action is largest at the origin, for constant, weak fields, and decreases away from that, faster in the direction of  $\kp$ than $g^3B/m^2$. Here the numerical, all-order results are shown in blue alongside, in dashed red, the analytic approximation containing only the leading-order correction due to self-interactions, Eq.~\eqref{eq:fullProbability}. Their close agreements shows that higher order corrections are small in this region of parameter space. In the top right, where the numerical results are absent, we were unable to obtain numerical solutions to the instanton equations due to the breakdown of Eq.~\eqref{eq:scale_hierarchy}.
}
  \label{fig:s_numerics}
\end{figure}

The blue curve in Fig.~\ref{fig:action_slice} shows the resulting instanton action for $g^3B/m^2=1$. One can see that the agreement with the leading-order corrected result~\eqref{eq:fullProbability} is good, and the full action appears to be slighly lower.
We were not able to reach higher $\xi$, where the higher-order corrections are expected to become more important, as for large large $\kp$ the worldlines become highly curved and it was not possible to maintain the necessary hierarchy of scales, Eq.~\eqref{eq:scale_hierarchy}.

Fig.~\ref{fig:s_numerics} shows the full action in the parameter region $(g^3B/m^2,\kp)\in ([0,1],[0,2.5])$. For the reasons discussed above, we were not able to obtain results for the top right corner of the plot.
We leave the numerical investigation of larger $g^3B/m^2$ and $\kp$ for future work.

Our numerical results show remarkably good agreement with Eq.~\eqref{eq:fullProbability}. Thus, at least in the regime we have considered, higher order terms in $g^3B/m^2$ are small. This might have been expected, given that all higher order terms in $g^3B/m^2$ vanish at $\kp=0$ \cite{affleck1981pair}. However, extrapolating the $O(g^3B/m^2)$ corrections to large $\kp$, one sees that they eventually dominate over the leading order term, making the action negative. Thus it is clear that higher order corrections must become important for large $\kp$.

\begin{figure}
 \centering
  \includegraphics[width=0.45\textwidth]{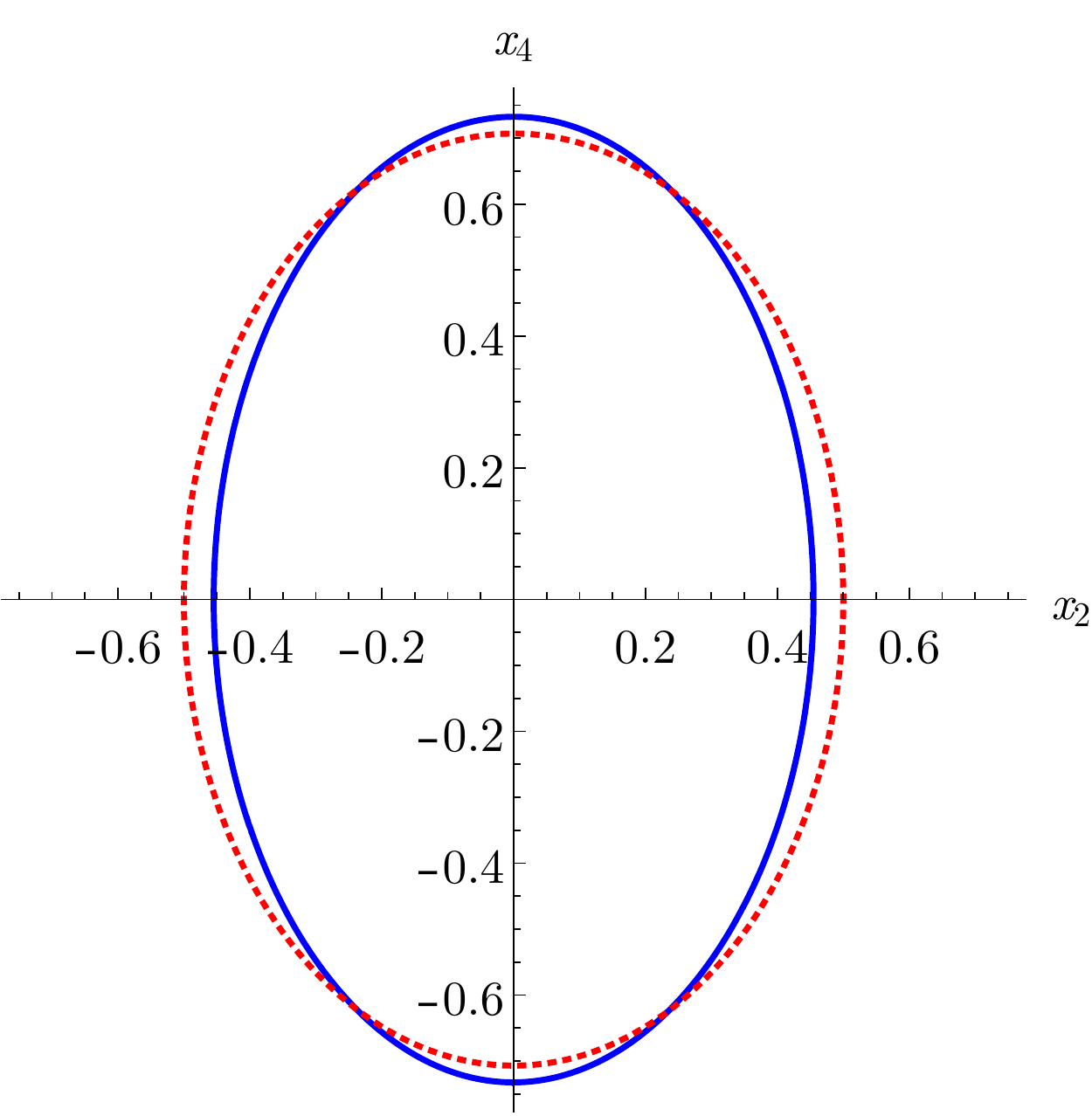}
  \caption{The worldline instanton to all orders in the self-interactions at $(g^3B/m^2,\kp)=(1,1)$, shown in blue. This is compared to the analytic result without self-interactions, i.e. at $(g^3B/m^2,\kp)=(0,1)$, in dashed red. Self-interactions give a modest increase to the maximum curvature of the worldline instanton.
}
  \label{fig:worldline_numerics}
\end{figure}

In Fig.~\ref{fig:worldline_numerics} we also show the effect of interactions on the shape of the worldline instanton. In the region of parameter space we have been able to explore numerically, interactions lead to a modest increase in the curvature of the worldline instanton. As we will discuss in Section \ref{sec:pointlike_approx}, this suggests that self-interactions do not prevent the breakdown of the small monopole approximation at large $\kp$.

\section{Consequences for monopole searches} \label{sec:consequences}
\subsection{Temporal `Inhomogeneity' as a property of the monopole}
The key result from Section \ref{sec:worldlineInstanton} is Eq.~\eqref{eq:fullProbability}, the exponential dependence of the monopole pair production probability. The  time dependence of the field of the heavy-ion collision enters the rate, and the corresponding worldline instanton, through a single dimensionless parameter $\xi$, defined in Eq.~(\ref{equ:xidef}) in terms of the peak value $B$ of the magnetic field and the decay constant $\omega$ of the field's time dependence. As discussed in Section \ref{sec:em_fields}, for peripheral collisions (the type most likely to produce monopoles), Eqs.~(\ref{eq:Bmaxgamma}) and (\ref{eq:omegamaxgamma}) give
\begin{align}
    B &\approx c_B \frac{Z e v \gamma}{2 \pi R^2} \label{eq:HIField}, \\
    \omega &\approx c_\omega \frac{v \gamma}{R} \label{eq:HIFreq},
\end{align}
where \(Z e\) is the heavy-ion charge, \(R\) is the heavy-ion radius (in its rest frame), \(v\) is the ion speed, \(\gamma = 1/\sqrt{1 - v^2}\) is the Lorentz factor of the collision, and \(c_B\) and \(c_\omega\) are \(O(1)\) dimensionless constants. It follows then that the temporal `inhomogeneity' of the magnetic field in a peripheral heavy-ion collision is given by
\begin{equation} \label{eq:HIKeldyshParam1}
    \kp \approx \frac{c_\omega}{c_B}\frac{2 \pi m R}{Z e g}.
\end{equation}
The most striking consequence of this observation is that the temporal inhomogeneity of the field is \emph{independent} of the energy of the collision. This may be understood by considering that, while the temporal extent of the field decreases proportionally to \(\gamma\), the increase in peak field strength causes a contraction of the worldline instanton that precisely cancels this effect. If the field `looks constant' --- i.e. does not vary significantly over the worldline instanton --- at any given relativistic energy, this holds for all relativistic energies.

The Keldysh parameter \eqref{eq:HIKeldyshParam1} for heavy-ion collisions can be expressed in an alternate form by utilising the Dirac quantisation condition \eqref{equ:Diraccondition}. This gives
\begin{equation}
    \kp \approx \frac{c_\omega}{c_B} \frac{m R}{Z n},
\end{equation}
where $n$ is the Dirac charge of the monopole.
The values of \(R\) and \(Z\) are specific to the colliding species, so for a given heavy-ion collision, \(\kp\) is proportional to the ratio of the monopole mass to the Dirac charge. Using the commonly accepted values for lead-lead collisions at the LHC of \(R = 6.62 \, \mr{fm}\), \(Z = 82\) \cite{alice2014centrality}, and the numerical fits \(c_\omega = 0.78\) and \(c_B = 0.92\) obtained in Section \ref{sec:em_fields}, 
\begin{equation} \label{eq:massScale}
    \kp \sim  \frac{m}{3n \, \mr{GeV}}.
\end{equation}
This suggests that, when considering production of monopoles with mass greater than \(\sim 3 \, \mr{GeV}\), the time dependence of the magnetic field cannot be neglected at any relativistic energy. The current best theoretical mass bounds \cite{gould2018mass} are close to this scale, and many theoretical monopoles (e.g. \cite{thooft1974magnetic,polyakov1974particle}) predict masses far greater. As a result, we conclude that the effects of time dependence are crucial to our understanding of potential magnetic monopole production in heavy-ion collisions.

For heavy monopoles (such that \(\kp \gg 1\)), the pair production probability has exponential dependence (to leading order in \((mR)^{-1}\))
\begin{equation}
	\ln P \sim -\frac{4 m R}{v \gamma} + \frac{\pi^2 n m R}{2 Z e^2},
	\label{eq:log_probability_heavy}
\end{equation}
where we have dropped the dependence on the $O(1)$ constants, $c_B$ and $c_\omega$, for simplicity.
Combining this with Eq.~\eqref{eq:approxPrefactor} for an approximation to the prefactor gives, for production of high-mass monopoles in peripheral heavy-ion collisions, 
\begin{equation}
	\frac{d\sigma}{db}\bigg|_{2R} \sim \frac{2 (2s+1) v^3\gamma ^2 n^4 Z^4}{9 \pi ^2 m^4 R^3}\mr{e}^{-\frac{4 m R}{v \gamma} + \frac{\pi^2 n m R}{2 Z e^2} } \label{eq:cross_section_heavy}
\end{equation}
up to an $O(1)$ multiplicative factor. The total cross section can be obtained by including the impact parameter dependence of the fields (see Section \ref{sec:em_fields}) and then integrating over all values of the impact parameter.

\subsection{Limitations of current approximations}
\label{sec:inapplicability}
The properties of heavy ions and the form of the magnetic fields in peripheral collisions are fixed, and, along with the Dirac quantisation condition, strongly constrain the parameter space in which our results could be applied. The only free parameters are the monopole mass, \(m\), its Dirac charge \(n\), and the collision Lorentz factor \(\gamma\), which for the LHC heavy ions is given in Eq.~(\ref{equ:gammaLHC}). In this section we examine the assumptions made in Section \ref{sec:worldlineInstanton} and show that there is unfortunately no region in this parameter space where all our approximations are valid.

\subsubsection{The semiclassical approximation}
The results of Section \ref{sec:worldlineInstanton} are valid providing that the conditions in Eqs.~\eqref{eq:semiclassical_condition} and \eqref{eq:curvature_condition} are met. The first of these is the semiclassical approximation, requiring the stationary value of the action to be large.

For high-mass monopoles, \(m \gg 3 n \, \mr{GeV}\),
the next-to-leading order action~(\ref{eq:log_probability_heavy}) is proportional to $mR\gg 1$, so the semiclassical approximation is satisfied as long as the action is positive (ignoring the fine-tuned edge case). As a result the semiclassicality condition is
\begin{equation}
    n v \gamma \lesssim \frac{8Z e^2}{\pi^2},
\end{equation}
or, taking $Z=82$ for lead,
\begin{equation} \label{eq:semiclassicalCondition}
    n v \gamma \lesssim 6.
\end{equation}
This condition is not satisfied in the LHC heavy ion collisions, because of their high Lorentz factor (\ref{equ:gammaLHC}).

The breakdown of the semiclassical approximation usually indicates unsuppressed particle production, as long as all other approximations are under control at this point. However, in our case (\ref{eq:log_probability_heavy}) it happens because the self-interaction correction becomes comparable to the tree-level action and cancels it. Therefore it merely shows that one needs to include the self-interaction to all orders, as was done in Section~\ref{sec:numerics}. However, in that section we were not able to explore the relevant regime, due to the difficulty of resolving the large hierarchy of scales that arises in this case.

While our current work focuses on magnetic monopoles, the need to include all orders in worldline self-interactions at high inhomogeneities is also relevant when considering Schwinger production of electrons. For high values of the Keldysh parameter, the curvature of the worldline instanton (scaled to its size) is so large that self-interactions cannot be ignored even for weak coupling. This explains the apparent `weak-field' divergence of the results in Ref.~\cite{lan2018holographic}: it in fact corresponds to a departure from the small self-interaction regime. Under such conditions, the non-self-interacting worldline solution is no longer a good approximation to the true saddlepoint solution of the full action. Increasing curvature with increasing temporal inhomogeneity appears to be a general feature of time-dependent fields \cite{dunne2005worldline}, so our current calculations and planned numerical work are relevant to a wider class of Schwinger production scenarios.

\subsubsection{Monopole size}
\label{sec:pointlike_approx}
\begin{figure}
 \centering
  \includegraphics[width=0.45\textwidth]{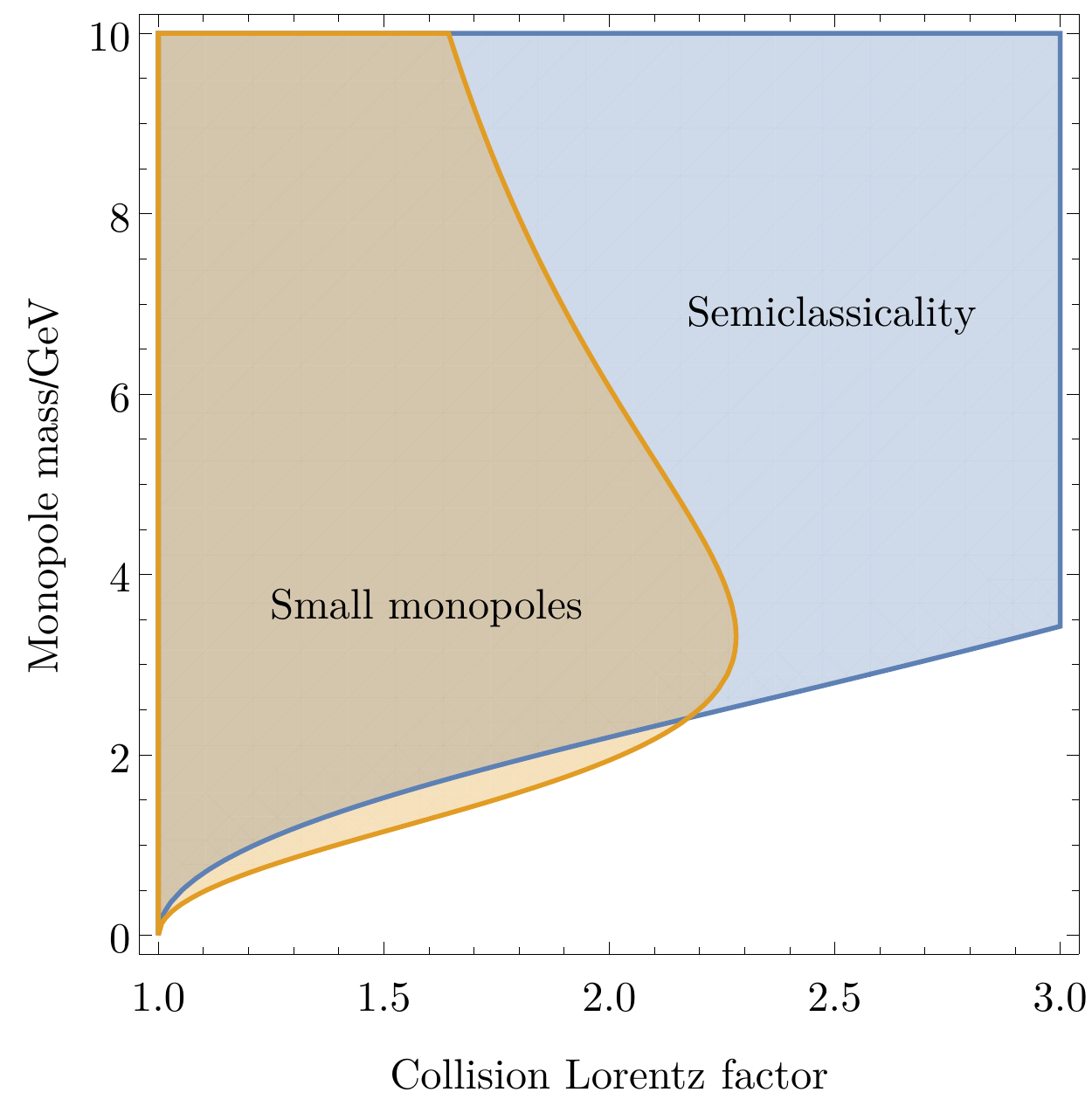}
  \caption{Plot showing the regions in the \(\gamma\)-\(m\) plane in which the approximations of semiclassicality (blue) and small monopoles (orange) are valid (assuming the Dirac charge \(n = 1\)). Note that in there is a turning point in the region of applicability of the small monopoles approximation, preventing us from going to large Lorentz factors.}
  \label{fig:applicabilityPlot}
\end{figure}
The second approximation our calculation relies upon is Eq.~\eqref{eq:curvature_condition}: the condition that the monopole size is small compared to the radius of curvature of the worldline instantons, validating our use of the worldline description (discussed in Section \ref{sec:general_approach}). Using  the radius of curvature of the ellipse \eqref{eq:noninteractingInstanton}, Eq.~\eqref{eq:curvature_condition} becomes
\begin{equation} \label{eq:point-likeCondition}
    \frac{2 m R v \gamma}{\pi Z^2 e^2} \ll 1.
\end{equation}
Assuming that the monopole mass is high, this is the most stringent constraint, requiring (for lead-lead collisions)
\begin{equation}
    m v \gamma \ll 10 \, \mr{GeV}.
\end{equation}
This limit prevents application of our results to any energies relevant to modern heavy-ion collisions, and for the energies at which the small-monopole approximation does apply, the Lorentz factor is too low to justify the assumptions (from the fits in Section \ref{sec:em_fields}) that the electromagnetic field varies more slowly in the transverse directions. As a result, we are unable at present to provide a reliable monopole production cross section.

For 't Hooft-Polyakov monopoles, one can overcome the limitations of the small-monopole approximation by performing an instanton calculation in the full field theory describing the monopole of interest. Such a calculation could be performed numerically using classical lattice field theory techniques.
For elementary monopoles, the effective monopole size arises from quantum effects, and therefore including it would require a non-perturbative quantum field theory calculation.

The inapplicability of our results to realistic heavy ion collisions at present is shown clearly in Fig.~\ref{fig:applicabilityPlot}. This shows the regions in the \(\gamma\)-\(m\) plane in which the small-monopole and small self-interaction approximations respectively hold, assuming monopoles with Dirac charge \(n = 1\). The boundary of the region in which the small-monopole assumption is valid has a turning point meaning that to probe \(\gamma \gtrsim 2\) (which is necessary if we are to apply the fits from Section \ref{sec:em_fields}) we must move beyond the worldline method.

Fig.~\ref{fig:applicabilityPlot} shows that the region in which the small-monopole approximation applies lies almost entirely within the region where the effect of worldline self-interactions are small. This suggests that moving beyond the small-monopole approximation is of the highest priority. The results of Section \ref{sec:numerics} showed that, at least in the region of parameter space we were able to study, self-interactions yield worldline instantons with somewhat higher curvature. This implies that the small-monopole approximation breaks down slightly earlier than suggested in Fig.~\ref{fig:applicabilityPlot}.

\section{Conclusions}\label{sec:conclusions}

\begin{figure}
 \centering
  \includegraphics[width=0.45\textwidth]{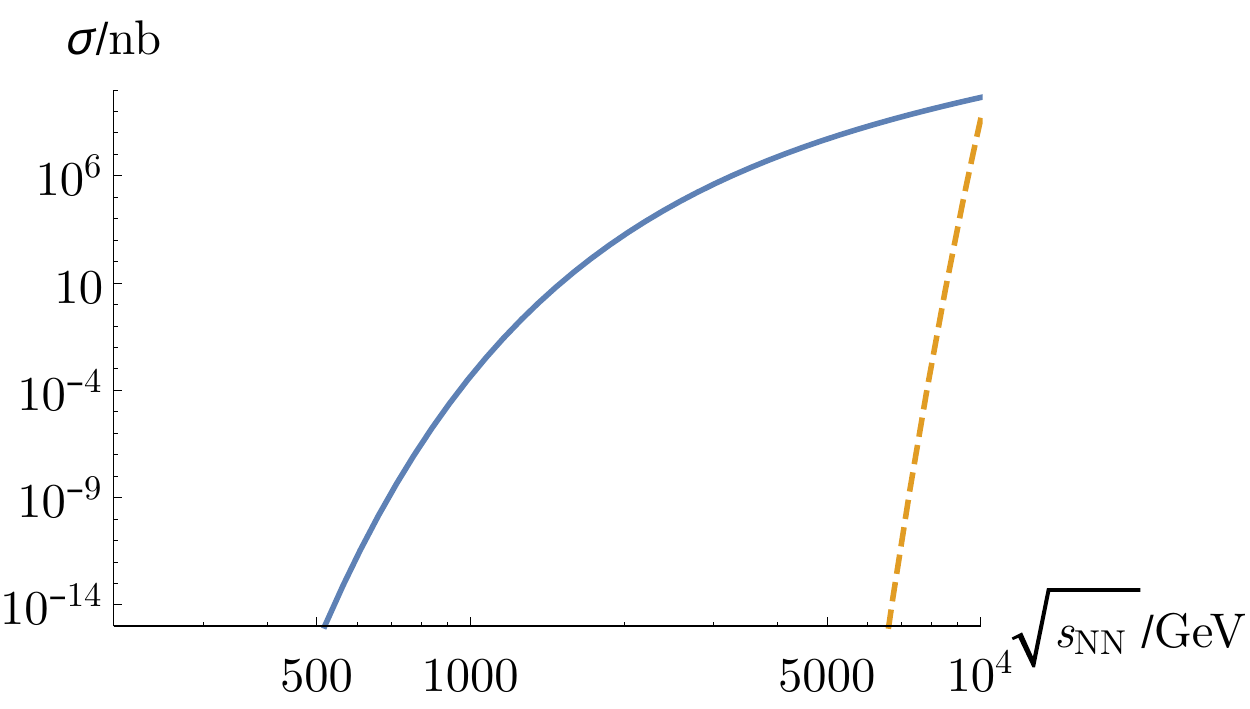}
  \caption{The total cross section in heavy-ion collisions for Schwinger production of magnetic monopoles with mass $m=$100~GeV, Dirac charge $n=1$ and spin $s=0$ is shown here in blue. The huge enhancements from the time-dependence can be seen by comparison with the locally-constant field result (see Appendix \ref{appendix:lcfa}) in dashed orange. Corrections from self-interactions (not shown here) provide even greater enhancements still. However, they also show the breakdown of our approximations at all relevant energies (see Section \ref{sec:inapplicability}).}
  \label{fig:crossSectionPlot}
\end{figure}

In this paper, we have computed the cross section for magnetic monopole production in ultrarelativistic heavy-ion collisions.
Our results hold for collision and monopole parameters such that the worldline instanton curvature is large compared to the size of the monopole --- this unfortunately removes the possibility of applying our results directly to real heavy-ion collisions at the LHC.

We have shown that, for Schwinger production of magnetic monopoles, the only relevant spacetime inhomogeneities in the electromagnetic field are time dependence and spatial variation along the beam axis, perpendicular to the direction in which the field points. In the worldline formalism, this feature of the spatial dependence means that we do not find the exponential suppression present in systems where there is significant spatial variation along the direction of the field (such as in the electric field in a heavy-ion collision, relevant for electron-positron pair production). The temporal inhomogeneity, which is the only relevant variation for computation of the worldline instanton, is well-approximated by an expression for which the exponential dependence of the pair production probability is known in closed form in the weak-field limit.

We have extended previous work on Schwinger pair production in spacetime dependent fields \cite{brezin1970pair,Nikishov:1970br,Popov:1971ff,popov1972pair,Marinov1977electron,dunne2005worldline,dunne2006worldline,kim2006schwinger} by including the dynamical effects of the photon field, which manifest as self-interactions in the worldline formalism. These are important for application to strongly-coupled monopoles. For the monopole production cross section, we find
\begin{align}
\sigma &\propto \exp \Bigg[-\frac{\pi m^2}{g B}\frac{4 [\mb{E}(-\kp^2) - \mb{K}(-\kp^2)]}{\pi \kp^2} \nonumber \\
&\qquad\quad +\frac{g^2}{8}\left(\sqrt{1 + \kp^2} + \frac{1}{\sqrt{1 + \kp^2}}\right)\Bigg].
\end{align}
Substituting parameter values relevant to heavy-ion collisions results in Eq.~\eqref{eq:cross_section_heavy}.

Our result, in agreement with previous analyses, encodes the temporal inhomogeneity of the field via a single dimensionless parameter 
$\xi$, defined by Eq.~(\ref{equ:xidef}), which depends on the mass $m$ and charge $g$ of the monopole, the peak strength $B$ of the magnetic field, and the field's decay constant $\omega$. As expected, the well-known constant field result \cite{affleck1981monopole,affleck1981pair} is obtained in the \(\kp \to 0\) limit.

As well as the properties of the monopoles, the total cross section depends on three collision parameters: the heavy-ion charge \(Z\), its radius \(R\), and the collision Lorentz factor \(\gamma\). Both \(B\) and \(\omega\) are linearly proportional to the ions' Lorentz factor, \(\gamma\), in the centre of mass frame (Eqs.~\eqref{eq:Bmaxgamma} and \eqref{eq:omegamaxgamma}). This means that \(\kp\) is independent of collision energy, and for a fixed ion species can be considered to be solely a property of the monopole. 
For collisions of lead ions, the condition for the time dependence of the field to have a significant effect is
\begin{equation}
m \gtrsim 3 n \, \mr{GeV}
\end{equation}
where the integer \(n\) is the number of Dirac charge quanta the monopole carries. If this condition is satisfied, taking the effects of time dependence into account is crucial at all relativistic collision energies.

Our results show that, when our approximations are valid, the time dependence of the collision and the effects of the monopole self-interactions both enhance the production rate compared with the constant-field Schwinger process. This suggests that if the calculation can be extended to realistic LHC heavy ion collisions, the mass bounds obtained would be stronger than previously estimated (see Fig. \ref{fig:crossSectionPlot}).
At least for 't Hooft-Polyakov monopoles, this can be done by performing an instanton calculation in the full field theory describing the monopole of interest. This would provide theoretically sound predictions for collider searches such as at the LHC.

\section*{Acknowledgements}
The authors wish to acknowledge CSC – IT Center for Science, Finland, for computational resources. O.G. was supported from the Research Funds of the University of Helsinki, D.H. was supported by a U.K. Science and Technology Facilities Council studentship and A.R. was supported by the U.K. Science and Technology Facilities Council grant ST/P000762/1.

\appendix

\section{Finite difference formulation\label{appendix:discretisation}} 

In this appendix we give our discrete approximation to the action in Eq.~\eqref{eq:action} and the corresponding equations of motion. We first integrate out the Schwinger parameter, $s$, and scale the worldlines by $gB/m$, making them dimensionless. We then discretise the worldline into $N$ points and use a simple finite difference approximation
\begin{align}
\frac{gB}{m^2}S[x]&=\sqrt{N\sum_{i,\mu} (x_\mu^{i+1}-x_\mu^i)^2} \nonumber \\
&+ \sum_i \frac{x_4^i}{\sqrt{1-(\kp x_4^i)^2}} (x_2^{i+1}-x_2^i)\nonumber \\
-\frac{g^3B}{2m^2}& \sum_{i,j}(x_\mu^{i+1}-x_\mu^i)(x_\mu^{j+1}-x_\mu^j)G_R(x^i,x^j;a) \label{eq:action_finite_difference}
\end{align}
where $i$ and $j$ run over $0,1,...,N-1$ and contractions of Euclidean indices $\mu$ are implied. As discussed in Ref.~\cite{gould2017thermal} we choose an exponential counterterm, rather than the simpler length counterterm of Polyakov \cite{polyakov1980gauge}, so that the bare mass is positive,
\begin{align}
G_R(x,y;a)&=\frac{-1}{4\pi^2((x-y)^2+a^2))} \nonumber \\
&\ +\frac{\sqrt{\pi}}{4\pi^2a^2}\mathrm{e}^{-(x-y)^2/a^2}.
\end{align}
This exponential counterterm is equivalent to the length counterterm for $a\to 0$, via the delta function limit of the Gaussian. Note also that we have dropped the gauge-dependent parts of $G_R$ as they integrate to zero over any closed worldline.

In the absence of spacetime inhomogeneity, there are translational zero modes which must be fixed to find a unique solution. We do this by imposing that the centre of mass of the worldline is at the origin. The timelike inhomogeneity of the background field breaks the translational symmetry in the time direction, though not that in the $x_2$ direction. There is also a zero mode resulting from the reparametrisation symmetry. We fix this by imposing $x_2^0-x_2^{N/2 - 1}=0$, which essentially fixes the point $i=0$ and $i=N/2-1$ to be at the bottom and top of the worldline. We fix all these constraints using Lagrange multipliers, $\lambda_\mu$ and $\sigma$, which amounts to,
\begin{equation}
\frac{gB}{m^2}S[x] \to \frac{gB}{m^2}S[x] + \sum_{i,\mu}\lambda_\mu x^i_\mu + \sigma (x_2^0-x_2^{N/2 - 1}).
\end{equation}
We then solve the $2N+2+1$ equations of motion derived from this action,
\begin{align}
\frac{\partial S}{\partial x_\mu^j} &= 0, \nonumber \\
\frac{\partial S}{\partial \lambda_\mu} &= 0, \nonumber \\
\frac{\partial S}{\partial \sigma} &= 0,
\end{align}
for $\mu=2,4$, $j=0,\dots,N-1$, using the Newton-Raphson method. The analytic solutions at $g^3B/m^2=0$ and at $\kp=0$ provide initial guesses for the solutions at small $g^3B/m^2$ and $\kp$ respectively. One can then step out in parameter space, using the solution at the previous parameter point as the initial guess. In this way we were able to solve the worldline instanton equations in the region shown in Fig.~\ref{fig:s_numerics}.

\section{The locally constant field approximation\label{appendix:lcfa}}

The locally constant field approximation (LCFA) is applicable when the electromagnetic fields vary on larger length and time scales than those of the worldline instanton, and hence the constant field result for pair production can be used locally. In this approximation, and when $F^{\mu\nu}\tilde{F}_{\mu\nu}=0$ (see Fig. \ref{fig:invariants}), the probability for monopole pair production is given by \cite{affleck1981monopole,Kohlfurst:2017git}
\begin{align}
    P_{\mr{LCFA}} &= \frac{(2s+1)g^2}{8\pi^3}\nonumber \\
    &\ \int \mathrm{d}^4x (B^2(x)-E^2(x))^2\mathrm{e}^{-\frac{\pi m^2}{g\sqrt{B^2(x)-E^2(x)}}+\frac{g^2}{4}},
\end{align}
Evaluating this for the fields of Eqs.~\eqref{eq:fit_functions}, and using the saddle point approximation for the integrals, gives
\begin{align}
    P_{\mr{LCFA}} &\approx \frac{(2s+1)(g B)^4}{18 \pi ^3 m^4 \omega^2 \Omega^2}\mathrm{e}^{-\frac{\pi m^2}{gB}+\frac{g^2}{4}},
\end{align}
where $\Omega$ is the decay rate of the field in the $x^1$ and $x^2$ directions and this formula is accurate up to an $O(1)$ multiplicative factor. Finally, using the impact parameter dependence of $B$ and $\omega$ (see Section \ref{sec:em_fields}) and integrating over the impact parameter, we arrive at the total cross section,
\begin{align}
    \sigma_{\mr{LCFA}} &\approx 9\times 10^{-3}\frac{\gamma ^{5/2} (n Z)^{9/2} }{m^5 R^3}\mr{e}^{-\frac{4.03 m^2 R^2}{\gamma v  n Z}+\frac{\pi ^2 n^2}{e^2}},
\end{align}
measured in $\mr{GeV}^{-2}$. This result is what is plotted in Fig. \ref{fig:crossSectionPlot}.

\bibliography{refs}

\end{document}